\documentclass[preprint,12pt]{elsarticle}

\usepackage{amsmath,amsthm,amssymb,bm,color,graphicx}
\usepackage{amscd,verbatim}
\usepackage[breaklinks=true,colorlinks=true,linkcolor=blue,urlcolor=blue,citecolor=blue]{hyperref}
\usepackage{graphicx}
\usepackage{subfigure}
\usepackage{comment}
\usepackage{lipsum}
\usepackage[arrow,matrix,curve]{xy}
\usepackage[numbers]{natbib}

\usepackage{tikz-cd}
\usetikzlibrary{decorations.markings}
\tikzset{->-/.style={decoration={
  markings,
  mark=at position #1 with {\arrow{>}}},postaction={decorate}}}
\tikzset{->-/.default=0.5}

\newcommand\wt{\widetilde}

\newcommand{\PP}{{\mathbb P}}
\newcommand{\RR}{{\mathbb R}}
\newcommand{\TT}{{\mathbb T}}
\newcommand{\ZZ}{{\mathbb Z}}

\newcommand{\vect}[1]{\mathbf{#1}}
\newcommand{\Tinv}{\mathsf{T}\text{-stable}}
\newcommand{\Ttorus}{\mathsf{T}\text{-plane}}
\newcommand{\Ttori}{\mathsf{T}\text{-planes}}
\newcommand{\Twsm}{\mathsf{T}\text{-WSM}}

\newcommand{\Tsurfs}{\mathsf{T}\text{-surfaces}}

\newcommand{\oeq}{\mathrel{\text{\textcircled{$=$}}}}

\journal{Nuclear Physics B}

\begin{document}

\begin{frontmatter}

\title{Fu--Kane--Mele monopoles in semimetals}

\author[1]{Guo Chuan Thiang}
\address[1]{School of Mathematical Sciences, University of Adelaide, SA 5000, Australia}
\author[2]{Koji Sato}
\address[2]{Institute for Materials Research, Tohoku University, Sendai 980-8577, Japan}
\author[3]{Kiyonori Gomi}
\address[3]{Department of Mathematical Sciences, Shinshu University, Matsumoto, Nagano 390-8621, Japan}




\begin{abstract}
In semimetals with time-reversal symmetry, the interplay between Weyl points and Fu--Kane--Mele indices results in coexisting surface Dirac cones and Fermi arcs that are transmutable without a topological phase transition. We show that Weyl points act as a new type of monopole, and that their connectivity is essential for capturing the full topology of semimetals and their role as intermediaries of topological insulator transitions. The history of Weyl point creation-annihilation provides a simple and mathematically equivalent way to classify semimetals, and directly prefigures the surface state topology. We further predict the possibility of a topological Dirac cone on the interface between two Weyl semimetals.
\end{abstract}

\begin{keyword}
Topological phases \sep Weyl semimetals \sep Poincar\'{e} duality \sep equivariant homology \sep Fu--Kane--Mele invariants

\PACS 02.40.Re

\MSC 55N91 \sep 81Q70

\end{keyword}

\end{frontmatter}

\section{Introduction}\label{sec:intro}

The groundbreaking discoveries of topological phases of matter in the last decade have fueled intense research activities in theoretical and experimental condensed matter physics. Topological insulators and superconductors, whose spectra have energy gaps, have unusual properties arising from topological invariants depending on both symmetries and dimension \cite{Hasan,QZ}. More recently, topological \emph{semimetals} have received a lot of attention due to them hosting exotic fermions as quasiparticles. For \emph{Weyl} semimetals (WSM) with broken $\mathsf{T}$ (time-reversal) symmetry, proposals include pyrochlore iridates \cite{Wan,Yang}, ferromagnetic compounds \cite{XuG}, and multilayer structures of topological insulators with ferromagnetic order and normal insulators \cite{Burkov}; $\mathsf{T}$-symmetric WSM $(\Twsm)$ were also proposed in \cite{HB}.

3D WSMs have topologically protected band crossings \cite{Wan,Vafek}: a two-band Bloch Hamiltonian $H(k)=\vect{h}(k)\cdot \bm{\sigma},\, k\in\TT^3$ is specified by a 3-component vector field $\vect{h}$ over the Brillouin zone (BZ) $\TT^3$, with $\bm{\sigma}$ the vector of Pauli matrices. Since the spectrum is $\pm|\vect{h}(k)|$, band crossings occur at a set $W\subset\TT^3$ of \emph{Weyl points} where $\vect{h}=\vect{0}$. The \emph{Weyl charge} of $w\in W$ is the degree (2D winding number) of $\widehat{\vect{h}}=\vect{h}/|\vect{h}|$ restricted to a small 2-sphere $S^2_w$ around $w$, and equals the Chern number of the valence band restricted to $S^2_w$, so $w$ is a (Dirac) monopole for the band's Berry curvature $\mathcal{F}$. 

A charge-cancellation condition \cite{NN} means that a \emph{topological} WSM has at least two oppositely-charged Weyl points. It was predicted that \emph{Fermi arcs} of surface states would connect the projected Weyl points on the surface BZ \cite{Wan,Burkov}. A pair of Weyl points (monopole-antimonopole pair) can be imagined as being connected by a ``Dirac string'' inside the Brillouin \emph{torus}. Because the Brillouin torus is not simply-connected, there are topologically inequivalent ``Dirac strings'' connecting the same pair of Weyl points, each associated to a topologically distinct WSM bulk band structure and surface Fermi arc \cite{MT,MT1}.

The experimental discovery of Fermi arcs in $\Twsm$ \cite{Lv,Xu1,Zhang} naturally lead to a search for materials with \emph{both} Fermi arcs and Dirac cones; the latter are associated with $\mathsf{T}$-symmetric topological insulators (TI). Such systems have been studied on the interface of a WSM-TI heterostructure \cite{Gru} and in $\Twsm$ \cite{Lau,Sheng}. However, the topological index there and in general $\Twsm$s is not understood precisely. 

We solve this problem in this paper, by defining topological invariants carefully for $\Twsm$, and computing them to be $\ZZ_2^4\oplus\ZZ$ for the simplest $\Twsm$ (with the minimal number of Weyl points --- four). The generalisation to more complicated $\Twsm$ with a larger set of Weyl points is then straightforward, see Section \ref{section:morepairs}. Our analysis reveals how Weyl points and Fu--Kane--Mele invariants (FKMI) interact in a subtle way with dramatic consequences for the surface states. Dirac cones and Fermi arcs are not ``topologically independent'', but can transmute between each other without moving the Weyl points or changing the bulk invariants, c.f.\ the ``dangling Dirac cone'' found in \cite{Lau}. Furthermore, we show that Weyl points in $\Twsm$ act as a new type of ``Fu--Kane--Mele monopole'' sourcing a $\ZZ_2$-invariant. We also provide a new \emph{arc-representation} of $\Twsm$, which may be interpreted physically as a classification scheme based on the history of Weyl point creation/annihilation. The arc-representation directly exhibits TI phase transitions, and prefigures the surface Fermi arc-Dirac cone topology. In particular, the FKMI of the TI obtained from a Weyl point creation-annihilation process can be deduced directly from the topology of the closed arc representing the process, circumventing the need to probe the bulk band topology directly. This surprising result can be understood as a physical realization of a mathematical \emph{Poincar\'{e} duality}. Finally, we predict and provide numerical evidence that Fermi arcs can combine into a solitary surface Dirac cone on the interface between two $\Twsm$.

\subsection*{Outline}
The main physical results are presented in Sections \ref{sec:topping}, \ref{sec:histories}, \ref{sec:bbc} in an intuitive way so as to be accessible to a broad audience, with the underlying mathematical background provided in Section \ref{sec:math} for specialists and completeness. In Section \ref{sec:topping}, we review the FKMI for topological insulators, then generalize them to define topological indices for $\Twsm$ carefully. In particular, we explain how Weyl points can be understood as monopoles for the weak FKMI. In Section \ref{sec:histories}, we provide an intuitive alternative classification of WSM and $\Twsm$ using the history of Weyl point creation and annihilation, and show that it is mathematically equivalent to the first generalized FKMI approach. This second approach greatly simplifies the bulk-boundary analysis, which we extend to a study of interface states in Section \ref{sec:bbc}. Mathematical background on (co)homology $H^n, H_n$, definitions and calculations of their ``$\mathsf{T}$-symmetric versions'' $\mathcal{H}^n, \mathcal{H}_n$ using Mayer--Vietoris sequences, Poincar\'{e} duality, and the bulk-boundary correspondence, are given in Section \ref{sec:math}, while Section \ref{section:morepairs} analyses $\Twsm$ with an arbitrary number of pairs of Weyl points. Numerical methods for the analysis of the interface between two $\Twsm$ are provided in Section \ref{sec:numerics}.

\section{Topological indices in the presence of time-reversal symmetry}\label{sec:topping}
\subsection{Fu--Kane--Mele invariants (FKMI) for topological insulators (TI)}
We will coordinatise the Brillouin torus $\TT^d$ by $k\in [-\pi,\pi]^d$, with $-\pi$ identified with $\pi$ for each component $k_i$. Hamiltonians with $\mathsf{T}$-symmetry ($\mathsf{T}^2=-1$) obey $\mathsf{T}H(k)\mathsf{T}^{-1}=H(\tau(k))$ with $\tau(k)=-k$ the momentum-reversing involutive map. In $d$-dimensions, there are $2^d$ {\bf T}ime {\bf R}eversal {\bf I}nvariant {\bf M}omenta (TRIM) with $k=-k$. The basic 2D/3D TI has two valence bands by Kramers degeneracy at the TRIM. Although Bloch eigenstates $|u_a(k)\rangle, a=1,2$ are smoothly definable on the whole BZ (all Chern numbers can be shown to vanish due to $\mathsf{T}$-symmetry), the ``$\mathsf{T}$-gauge condition'' $|u_1(-k)\rangle=\mathsf{T}|u_2(k)\rangle$ may fail. From the matrix function $\omega(k)\equiv(\omega_{ab}(k))=\langle u_a(-k)|\mathsf{T}|u_b(k)\rangle$ measuring this failure, each TRIM may be assigned a $\pm$ sign given by the value of $\sqrt{\text{det}(\omega)}/\text{Pf}(\omega)$ there. These signs depend on the choice of $|u_a(k)\rangle$, but in 2D the product-of-signs (POS), called the 2D FKMI $\nu$, is \emph{gauge invariant} \cite{FK,FKM,dNG}; it also equals an {\bf E}ffective {\bf BZ} (EBZ) (e.g.\ take the front face of Fig.~\ref{fig:torsionloops}(a)) integral,
\begin{equation}
\nu=(-1)^\Delta,\;\;\Delta=\frac{1}{2\pi}\left(\oint_{\partial\rm EBZ}\mathcal{A}-\int_{\rm EBZ}\mathcal{F}\right)\,\,{\rm mod\,2}. \label{FKMIintegral}
\end{equation}
Note that the loop integral of the $\mathrm{U}(1)$ Berry connection $\mathcal{A}$ is computed in a $\mathsf{T}$-gauge over the EBZ boundary (which comprises two loops), and is gauge-invariant mod 2 (see \cite{FK} Appendix).

A \emph{$\Tinv$} subspace $Y$ has $\tau(y)\in Y$ whenever $y\in Y$. Examples include the $k_i=0$ or $\pi$, $i\in\{x,y,z\}$ planes in the 3D BZ, which we call \emph{$\Ttori$} (they are actually closed 2D tori). Applying $\nu$ to them gives six $\ZZ_2$-numbers $\nu_{i,0},\nu_{i,\pi}$. They obey $\nu_{i,0}=\nu_0\nu_{i,\pi}$, where $\nu_0$ is the \emph{strong} FKMI defined as the POS of all eight TRIM in the 3D BZ \cite{FKM}. Thus 3D TI are labelled by $(\nu_0;\nu_i)\in\ZZ_2^4$, with $\nu_i\equiv\nu_{i,\pi}$ the three \emph{weak} FKMI.

\subsection{Topological indices of $\mathsf{T}$-invariant Weyl semimetals ($\Twsm$)} 
For a valence band \emph{without} $\mathsf{T}$-symmetry, the Berry curvature 2-form $\frac{\mathcal{F}}{2\pi}$ integrates to an integer (a \emph{Chern number}) over \emph{any} closed surface in the BZ. In 3D, there are three independent (weak) Chern numbers $c_1^{xy},c_1^{yz},c_1^{zx}$ corresponding to the 2D subtorus at (any) fixed $k_z, k_x, k_y$ respectively. Valence bands of WSM are only defined over $\TT^3\setminus W$, and there is an extra Chern number on each $S^2_w$ corresponding to the Weyl charge of $w$. Then $c_1^{xy}$ depends on $k_z$, jumping by the Weyl charge when the constant $k_z$ subtorus is moved past $w$; similar jumps occur for $c_1^{yz},c_1^{zx}$. If a closed surface encloses zero net Weyl charge, the valence bands can be extended into the whole interior by annihilating oppositely-charged pairs of enclosed Weyl points and opening a gap. In reverse, Weyl point pairs can be created by gap closing and then separated, their ``history'' being a set of connections between $\pm$ Weyl pairs, oriented according to net positive charge motion (so the connections always point from $-$ to $+$). Even a single Weyl pair has several topologically distinct connections due to the periodic BZ directions, leading to topologically distinct WSM \cite{MT,MT1}. Each such connection is a topologically distinct ``Dirac string'' whose corresponding WSM has non-zero weak Chern numbers on those 2D subtori transverse to the string, see Section 3.1 and Fig.\ 5-6 of \cite{MT1}.

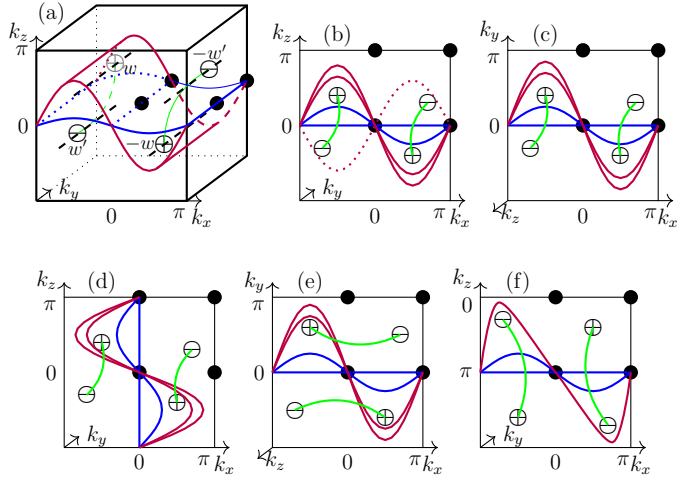
\begin{figure}[bth]
\centering
\subfigure{
\begin{tikzpicture}[xscale=1, every node/.style={scale=0.7}]
\draw[->](-0.4,1.7) -- (-0.4,1.9);
\draw[->](1.6,-0.3) -- (1.8,-0.3);
\draw[->](-0.4,-0.3) -- (-0.2,-0.15);

\node at (-0.2,2.2) {(a)};

\draw [thick] (-0.4,-0.3) -- (-0.4,1.7) -- (1.6,1.7) -- (1.6,-0.3) -- (-0.4,-0.3);
\draw [thick] (-0.4,1.7)  -- (0.4,2.3) -- (2.4,2.3) -- (2.4,0.3) -- (1.6,-0.3);
\draw [thick] (1.6,1.7) -- (2.4,2.3);

\node at (0.15,0.6) [scale=1.05] {$\boldsymbol{\ominus}$} ;
\node at (0.65,1.53) [gray,scale=1.05] {$\boldsymbol{\oplus}$} ;
\node at (1.3,0.46) [scale=1.05] {$\boldsymbol{\oplus}$} ;
\node at (1.9,1.45) [scale=1.05] {$\boldsymbol{\ominus}$} ;
\node [below] at (0.15,0.6) [scale=0.9] {$w'$} ;
\node [right] at (0.66,1.45) [scale=0.9] {$w$} ;
\node [left] at (1.27,0.46) [scale=0.9] {$-w$} ;
\node [above] at (1.9,1.47) [scale=0.9] {$-w'$} ;

\draw[green,dashed] (0.15,0.6) to [bend right] (0.65,1.53);
\draw[green] (1.9,1.45) to [bend right] (1.3,0.46);

\draw [thick,dashed] (-0.1,0.4) -- (0.7,1);
\draw [thick,dashed] (0.1,1.1) -- (0.9,1.7);
\draw [thick,dashed] (1.1,0.3) -- (1.9,0.9);
\draw [thick,dashed] (1.3,1) -- (2.1,1.6);

\draw [thin,dotted] (0.4,0.3) -- (0.4,2.3);
\draw [thin,dotted] (-0.4,-0.3) -- (0.4,0.3);
\draw [thin,dotted] (0.4,0.3) -- (2.4,0.3);

\node [left] at (-0.4,1.95) {$k_z$} ;
\node [below] at (1.8,-0.3) {$k_x$} ;
\node [right] at (-0.2,-0.15) {$k_y$};
\node [left] at (-0.4,1.7) {$\pi$} ;
\node [left] at (-0.4,0.7) {$0$} ;
\node [below] at (0.6,-0.3) {$0$} ;
\node [below] at (1.5,-0.3) {$\pi$} ;
\node at (1,1) [circle,fill=black,scale=0.7]{} ;
\node at (2,1) [circle,fill=black,scale=0.7]{} ;
\node at (1.4,1.3) [circle,fill=black,scale=0.7]{} ;
\node at (2.4,1.3) [circle,fill=black,scale=0.7]{} ;
\draw [blue,thick,dotted] (-0.4,0.7) -- (0.4,1.3);
\draw [blue,thick,dotted] (0.6,0.7) -- (1.4,1.3);

\draw [blue,thick] (1.6,0.7) -- (2.4,1.3);
\draw [purple,thick] (0,1.27) -- (0.8,1.87);
\draw [purple,thick] (1.2,0.13) -- (2,0.73);

\draw[thick,blue,domain=0:2] plot (\x-0.4, {0.7+0.1*sin(\x*pi r) });
\draw[thick,purple,domain=0:2] plot (\x-0.4, {0.7+0.6*sin(\x*pi r) });
\draw[blue,domain=1:2] plot (\x+0.4, {1.3+0.1*sin(\x*pi r) });
\draw[thick,blue,dotted,domain=0:1] plot (\x+0.4, {1.3+0.1*sin(\x*pi r) });
\draw[thick,dotted,purple,domain=0:0.4] plot (\x+0.4, {1.3+0.6*sin(\x*pi r) });
\draw[thick,purple,domain=0.4:1] plot (\x+0.4, {1.3+0.6*sin(\x*pi r) });
\draw[thick,dashed,purple,domain=1:2] plot (\x+0.4, {1.3+0.6*sin(\x*pi r) });

\end{tikzpicture}
}
\hspace{-1em}
\subfigure{
\begin{tikzpicture}[xscale=1, every node/.style={scale=0.7}]

\node at (0.5,2.2) {(b)};

\draw (0,0) -- (0,2) -- (2,2) -- (2,0) -- (0,0);
\draw[->] (0,2) -- (0,2.2);
\draw[->] (2,0) -- (2.2,0);
\draw[->] (0,0) -- (0.2,0.15);

\node [right] at (0.2,0.15) {$k_y$};
\node [left] at (0,2.23) {$k_z$} ;
\node [below] at (2.13,0) {$k_x$} ;
\node [below] at (1.86,0) {$\pi$} ;
\node [left] at (0,1.9) {$\pi$} ;
\node [left] at (0,1) {$0$};
\node [below] at (1,0) {$0$};

\node at (1,1) [circle,fill=black,scale=0.7]{} ;
\node at (2,1) [circle,fill=black,scale=0.7]{} ;
\node at (2,2) [circle,fill=black,scale=0.7]{} ;
\node at (1,2) [circle,fill=black,scale=0.7]{} ;

\node at (0.3,0.7) {$\boldsymbol{\ominus}$} ;
\node at (0.5,1.4) {$\boldsymbol{\oplus}$} ;
\node at (1.5,0.6) {$\boldsymbol{\oplus}$} ;
\node at (1.7,1.3) {$\boldsymbol{\ominus}$} ;

\draw[thick,blue,domain=0:2] plot (\x, {1+0.25*sin(\x*pi r) });
\draw[thick,blue,domain=0:2] plot (\x, 1);
\draw[thick,purple,dotted,domain=0:2] plot (\x, {1-0.6*sin(\x*pi r) });
\draw[thick,purple,domain=0:2] plot (\x, {1+0.8*sin(\x*pi r) });
\draw[thick,purple,domain=0:2] plot (\x, {1+0.65*sin(\x*pi r) });

\draw [green,thick] (0.3,0.7) to [bend right] (0.5,1.4);
\draw [green,thick] (1.7,1.3) to [bend right] (1.5,0.6);

\end{tikzpicture}
}
\hspace{-1.5em}
\subfigure{

\begin{tikzpicture}[xscale=1, every node/.style={scale=0.7}]

\node at (0.5,2.2) {(c)};

\draw (0,0) -- (0,2) -- (2,2) -- (2,0) -- (0,0);
\draw[->] (0,2) -- (0,2.2);
\draw[->] (2,0) -- (2.2,0);
\draw[->] (0,0) -- (-0.2,-0.15);

\node [right] at (-0.2,-0.17) {$k_z$};
\node [left] at (0,2.25) {$k_y$} ;
\node [below] at (2.13,0) {$k_x$} ;
\node [below] at (1.9,0) {$\pi$} ;
\node [left] at (0,1.9) {$\pi$} ;
\node[left] at (0,1) {$0$};
\node [below] at (1,0) {$0$};

\node at (1,1) [circle,fill=black,scale=0.7]{} ;
\node at (2,1) [circle,fill=black,scale=0.7]{} ;
\node at (2,2) [circle,fill=black,scale=0.7]{} ;
\node at (1,2) [circle,fill=black,scale=0.7]{} ;

\node at (0.3,0.7) {$\boldsymbol{\ominus}$} ;
\node at (0.5,1.4) {$\boldsymbol{\oplus}$} ;
\node at (1.5,0.6) {$\boldsymbol{\oplus}$} ;
\node at (1.7,1.3) {$\boldsymbol{\ominus}$} ;

\draw[thick,blue,domain=0:2] plot (\x,{1+0.25*sin(\x*pi r) });
\draw[thick,blue,domain=0:2] plot (\x,1);
\draw[thick,purple,domain=0:2] plot (\x,{1+0.85*sin(\x*pi r) });
\draw[thick,purple,domain=0:2] plot (\x,{1+0.7*sin(\x*pi r) });

\draw [green,thick] (0.3,0.7) to [bend right] (0.5,1.4);
\draw [green,thick] (1.7,1.3) to [bend right] (1.5,0.6);

\end{tikzpicture}
}

\subfigure{
\begin{tikzpicture}[scale=1, every node/.style={scale=0.7}]

\node at (0.5,2.2) {(d)};

\draw (0,0) -- (0,2) -- (2,2) -- (2,0) -- (0,0);
\draw[->] (0,2) -- (0,2.2);
\draw[->] (2,0) -- (2.2,0);
\draw[->] (0,0) -- (0.2,0.15);

\node [right] at (0.2,0.15) {$k_y$};
\node [left] at (0,2.23) {$k_z$} ;
\node [below] at (2.13,0) {$k_x$} ;
\node [below] at (1.86,0) {$\pi$} ;
\node [left] at (0,1.9) {$\pi$} ;
\node[left] at (0,1) {$0$};
\node [below] at (1,0) {$0$};

\node at (1,1) [circle,fill=black,scale=0.7]{} ;
\node at (2,1) [circle,fill=black,scale=0.7]{} ;
\node at (2,2) [circle,fill=black,scale=0.7]{} ;
\node at (1,2) [circle,fill=black,scale=0.7]{} ;

\node at (0.3,0.7) {$\boldsymbol{\ominus}$} ;
\node at (0.5,1.4) {$\boldsymbol{\oplus}$} ;
\node at (1.5,0.6) {$\boldsymbol{\oplus}$} ;
\node at (1.7,1.3) {$\boldsymbol{\ominus}$} ;

\draw[thick,blue,domain=0:2] plot ({1+0.3*sin(\x*pi r) },\x);
\draw[thick,blue,domain=0:2] plot (1,\x);
\draw[thick,purple,domain=0:2] plot ({1+0.85*sin(\x*pi r) },\x);
\draw[thick,purple,domain=0:2] plot ({1+0.7*sin(\x*pi r) },\x);

\draw [green,thick] (0.3,0.7) to [bend right] (0.5,1.4);
\draw [green,thick] (1.7,1.3) to [bend right] (1.5,0.6);

\end{tikzpicture}
}
\hspace{-1.5em}
\subfigure{
\begin{tikzpicture}[scale=1, every node/.style={scale=0.7}]

\node at (0.5,2.2) {(e)};

\draw (0,0) -- (0,2) -- (2,2) -- (2,0) -- (0,0);
\draw[->] (0,2) -- (0,2.2);
\draw[->] (2,0) -- (2.2,0);
\draw[->] (0,0) -- (-0.2,-0.15);

\node [right] at (-0.2,-0.17) {$k_z$};
\node [left] at (0,2.2) {$k_y$} ;
\node [below] at (2.13,0) {$k_x$} ;
\node [below] at (1.9,0) {$\pi$} ;
\node [left] at (0,1.9) {$\pi$} ;
\node[left] at (0,1) {$0$};
\node [below] at (1,0) {$0$};

\node at (1,1) [circle,fill=black,scale=0.7]{} ;
\node at (2,1) [circle,fill=black,scale=0.7]{} ;
\node at (2,2) [circle,fill=black,scale=0.7]{} ;
\node at (1,2) [circle,fill=black,scale=0.7]{} ;

\node at (0.3,0.5) {$\boldsymbol{\ominus}$} ;
\node at (0.5,1.6) {$\boldsymbol{\oplus}$} ;
\node at (1.5,0.4) {$\boldsymbol{\oplus}$} ;
\node at (1.7,1.5) {$\boldsymbol{\ominus}$} ;

\draw[thick,blue,domain=0:2] plot (\x,{1+0.25*sin(\x*pi r) });
\draw[thick,blue,domain=0:2] plot (\x,1);
\draw[thick,purple,domain=0:2] plot (\x,{1+0.9*sin(\x*pi r) });
\draw[thick,purple,domain=0:2] plot (\x,{1+0.75*sin(\x*pi r) });

\draw [green,thick] (0.3,0.5) to [bend left] (1.5,0.4);
\draw [green,thick] (1.7,1.5) to [bend left] (0.5,1.6);

\end{tikzpicture}
}
\hspace{-1.5em}
\subfigure{
\begin{tikzpicture}[scale=1, every node/.style={scale=0.7}]

\node at (0.5,2.2) {(f)};

\draw (0,0) -- (0,2) -- (2,2) -- (2,0) -- (0,0);
\draw[->] (0,2) -- (0,2.2);
\draw[->] (2,0) -- (2.2,0);
\draw[->] (0,0) -- (0.2,0.15);

\node [right] at (0.2,0.15) {$k_y$};
\node [left] at (0,2.25) {$k_z$} ;
\node [below] at (2.13,0) {$k_x$} ;
\node [below] at (1.9,0) {$\pi$} ;
\node [left] at (0,1.9) {$0$} ;
\node [left] at (0,1) {$\pi$};
\node [below] at (1,0) {$0$};

\node at (1,1) [circle,fill=black,scale=0.7]{} ;
\node at (2,1) [circle,fill=black,scale=0.7]{} ;
\node at (2,2) [circle,fill=black,scale=0.7]{} ;
\node at (1,2) [circle,fill=black,scale=0.7]{} ;

\node at (0.3,1.7) {$\boldsymbol{\ominus}$} ;
\node at (0.5,0.4) {$\boldsymbol{\oplus}$} ;
\node at (1.5,1.6) {$\boldsymbol{\oplus}$} ;
\node at (1.7,0.3) {$\boldsymbol{\ominus}$} ;

\draw[thick,blue,domain=0:2] plot (\x, {1+0.25*sin(\x*pi r) });
\draw[thick,blue,domain=0:2] plot (\x, 1);
\draw [thick,purple] plot [smooth] coordinates { (0,1) (0.25,1.93) (1,1) (1.75,0.07) (2,1)};

\draw [green,thick] (0.3,1.7) to [bend left] (0.5,0.4);
\draw [green,thick] (1.7,0.3) to [bend left] (1.5,1.6);

\end{tikzpicture}
}
\caption{(a) Purple/blue $\Ttori$ have the same EBZ boundaries. Their joint EBZ (left half planes) enclose $w$ and has Chern number $q$. The 2D FKMI jumps by $(-1)^q$ as a blue $\Ttorus$ is deformed across $\pm w$ into a purple one. (b)-(d) 2D BZ sections; green paths indicate how Weyl points are to be annihilated ``vertically''. (e) Horizontal annihilation leads to a different TI. (f) A different ``vertical'' annihilation path passing through $k_z=\pi$ instead of $k_z=0$.
}
\label{fig:familyoftori}

\end{figure}

\begin{figure}[bth]
    \centering
\subfigure{
\begin{tikzpicture}[scale=1, every node/.style={scale=0.7}]
\draw (0,0) -- (2,0) -- (2,2) -- (0,2) -- (0,0);
\draw [->] (0,2) -- (0,2.2);
\draw [->] (2,0) -- (2.2,0);
\draw [->] (0,0) -- (0.2,0.2);

\node [left] at (0,2.2) {$k_z$} ;
\node [above] at (2.2,0) {$k_x$} ;
\node [right] at (0.15, 0.15) {$k_y$};
\node [below] at (2,0) {$\pi$} ;
\node [left] at (0,1.9) {$\pi$} ;
\node [left] at (0,1) {$0$};
\node [below] at (1,0) {$0$};

\node at (1,1) [circle,fill=black]{} ;
\node at (2,1) [circle,fill=black]{} ;
\node at (2,2) [circle,fill=black]{} ;
\node at (1,2) [circle,fill=black]{} ;

\node at (1,1.707) {$\pm$};
\node at (1,0.293) {$\pm$};

\draw [dotted,->,domain=-0.1:-0.5] plot (1+\x, {1+(0.5-\x*\x)^(0.5) });
\draw [dotted,->,domain=0.1:0.5] plot (1+\x, {1+(0.5-\x*\x)^(0.5) });
\draw [dotted,->,domain=0:-0.5] plot (1+\x, {1-(0.5-\x*\x)^(0.5) });
\draw [dotted,->,domain=0:0.5] plot (1+\x, {1-(0.5-\x*\x)^(0.5) });

\draw (2.5,0) -- (4.5,0) -- (4.5,2) -- (2.5,2) -- (2.5,0);
\draw [->] (2.5,2) -- (2.5,2.2);
\draw [->] (4.5,0) -- (4.7,0);
\draw [->] (2.5,0) -- (2.7,0.2);

\draw [thick,brown] (2.7,0.3) -- (3.2,0.3) -- (3.2,1.7) -- (2.7,1.7) -- (2.7,0.3);
\draw [thick,brown] (3.8,0.3) -- (4.3,0.3) -- (4.3,1.7) -- (3.8,1.7) -- (3.8,0.3);
\draw [thin,cyan]  (2.8,0.2) -- (4.2,0.2) -- (4.2,0.7) -- (2.8,0.7) -- (2.8,0.2);
\draw [thin,cyan]  (2.8,1.3) -- (4.2,1.3) -- (4.2,1.8) -- (2.8,1.8) -- (2.8,1.3);

\node at (3.5,1) [circle,fill=black]{} ;
\node at (4.5,1) [circle,fill=black]{} ;
\node at (4.5,2) [circle,fill=black]{} ;
\node at (3.5,2) [circle,fill=black]{} ;

\node at (3,1.5) {$\boldsymbol{\oplus}$};
\node at (4,0.5) {$\boldsymbol{\oplus}$};
\node at (4,1.5) {$\boldsymbol{\ominus}$};
\node at (3,0.5) {$\boldsymbol{\ominus}$};

\draw [red,->-,domain=0.5:-0.5] plot (3.5+\x, {1+(0.5-\x*\x)^(0.5) });
\draw [red,->-,domain=-0.5:0.5] plot (3.5+\x, {1-(0.5-\x*\x)^(0.5) });

\draw [dotted,->,domain=-0.5:-0.7] plot (3.5+\x, {1+(0.5-\x*\x)^(0.5) });
\draw [dotted,->,domain=0.5:0.7] plot (3.5+\x, {1+(0.5-\x*\x)^(0.5) });
\draw [dotted,->,domain=-0.5:-0.7] plot (3.5+\x, {1-(0.5-\x*\x)^(0.5) });
\draw [dotted,->,domain=0.5:0.7] plot (3.5+\x, {1-(0.5-\x*\x)^(0.5) });

\draw (5,0) -- (7,0) -- (7,2) -- (5,2) -- (5,0);
\draw [->] (5,2) -- (5,2.2);
\draw [->] (7,0) -- (7.2,0);
\draw [->] (5,0) -- (5.2,0.2);

\node at (6,1) [circle,fill=black]{} ;
\node at (7,1) [circle,fill=black]{} ;
\node at (7,2) [circle,fill=black]{} ;
\node at (6,2) [circle,fill=black]{} ;

\node at (5.27,1) {$\pm$};
\node at (6.73,1) {$\pm$};

\draw [red,domain=-0.707:0.707] plot (6+\x, {1+(0.5-\x*\x)^(0.5) });
\draw [red,->-,domain=-0.707:0.707] plot (6+\x, {1-(0.5-\x*\x)^(0.5) });

\end{tikzpicture}
}

\caption{[L-R] Two $\pm$ pairs of Weyl points are created from a trivial TI, moved apart along the dotted arrows to form a $\Twsm$, then annihilated to form a strong TI. Their history traces out a circle (red) which represents the acquisition of a strong FKMI. The brown and cyan $\Tsurfs$ are two inequivalent ways to enclose cancelling pairs of Weyl points.}
\label{fig:ambiguity}
\end{figure}
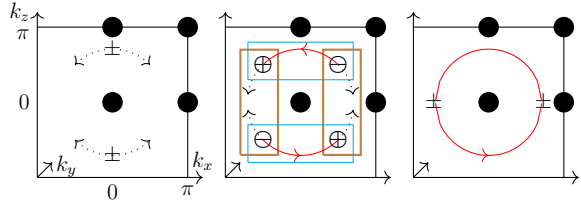

When there is $\mathsf{T}$-symmetry, a Weyl point $w$ must come with a corresponding one at $-w$ with the same Weyl charge. Thus the simplest $\Twsm$ has four distinct Weyl points $\pm w,\pm w'$, away from the TRIM and the $\Ttori$ at $k_i=0,\pi$ (Fig.~\ref{fig:familyoftori}(a)); $\pm w$ (resp.\ $\pm w'$) each has charge $q\in\ZZ$ (resp.\ $-q$) \cite{Kourtis,Lau,HB}. It has 2D FKMI $\nu_{i,0},\nu_{i,\pi}$ by restricting to respective $\Ttori$. However, the strong FKMI $\nu_0$ cannot be directly defined, because the Weyl charges prevent $|u_1(k)\rangle, |u_2(k)\rangle$ from existing globally on $\TT^3\setminus W$. To proceed, we first locally ``cordon off'' the Weyl points in charge-cancelling pairs by some $\mathsf{T}$-stable surfaces ($\Tsurfs$) $S$ that avoid the TRIM, e.g.\ the ``vertical'' brown surfaces in Fig.~\ref{fig:ambiguity} (center). Then $|u_a(k)\rangle$ can be globally defined \emph{on} and \emph{outside} of $S$ since all the Chern numbers vanish, so we can now calculate signs for all eight TRIM simultaneously. We then \emph{define} $\nu_0$ for the $\Twsm$ to be the resulting POS on the eight TRIM. With this definition, $\nu_0$ coincides with the strong FKMI $\nu_0^{\rm TI}$ for the resulting TI upon annihilation of the Weyl points within $S$. There is, however, a crucial ambiguity in pairing up the Weyl points, whose consequence is best illustrated by the trivial insulator $\rightarrow$ strong TI transition described in \cite{HB}, c.f.\ \cite{OM}. Starting from the intermediate $\Twsm$ (Fig.~\ref{fig:ambiguity}, center) and running the transition history backwards/forwards, we see that the two ways (horizontal/vertical) of annihilating the Weyl points result in TIs whose strong FKMI $\nu_0^{\rm TI}$ differ! We proceed with the ``vertical'' pairing convention, and resolve this $\nu_0$ ambiguity afterwards (Section \ref{sec:histories}).


In \cite{Gru,Lau}, it was found that the TI relation $\nu_{i,0}^{\rm TI}=\nu_0^{\rm TI}\nu_{i,\pi}^{\rm TI}$ no longer holds for $\Twsm$, suggesting that all six numbers $\nu_{i,0},\nu_{i,\pi}$ may become ``independent'' in a $\Twsm$. In fact, only \emph{three} are independent for a given $W$, the precise relation for the Fig.~\ref{fig:familyoftori}(a) example being
\begin{equation}
\nu_{i,0}=\begin{cases} (-1)^q\nu_0\nu_{i,\pi},\quad i=y,z,\\ \nu_0\nu_{i,\pi},\qquad\quad\;\, i=x.\end{cases} \label{signrelation}
\end{equation}
The extra $(-1)^q$ factors in Eq.\ \eqref{signrelation} may be understood by thinking of Weyl points as ``monopoles'' for the 2D FKMI invariant. Observe that there is a \emph{continuum} of \emph{curved} $\Ttori$ besides the standard flat ones (Fig.~\ref{fig:familyoftori}). Each curved $\Ttorus$ with no Weyl points also has its own 2D FKMI, stable under $\Ttorus$ deformation. For each $i=x,y,z$, consider a blue and a purple $\Ttorus$ whose EBZs share the \emph{same} boundary and jointly enclose $w$ (Fig.~\ref{fig:familyoftori}). From Eq.\ \eqref{FKMIintegral}, their respective $\Delta$ satisfy
\begin{equation}
\Delta_{\rm purple}^{i,0}-\Delta_{\rm blue}^{i,0}=\int_{{\rm purple\, EBZ} \atop -{\rm blue\, EBZ}}\mathcal{F}=\int_{{\rm joint} \atop {\rm EBZ} }\mathcal{F}=q\,\,{\rm (mod\,2)},\nonumber
\end{equation}
where the loop integral contributions cancel and the joint EBZ is oriented with outward normal\footnote{The valence bands cannot generally be trivialised over both blue and purple EBZs \emph{simultaneously} due to the enclosed Weyl point. TRIM signs calculated for the blue $\Ttorus$ cannot also be applied to the purple one, and different POS may be obtained in each case.}. It follows that 
\begin{equation}
\nu_{i,0}=\nu_{i,0}^{\rm \small blue}=(-1)^{\Delta_{\rm blue}^{i,0}}=(-1)^{\Delta_{\rm purple}^{i,0}-q}=(-1)^{q}\nu_{i,0}^{\rm purple}.\label{blue-purple}
\end{equation}
\normalsize
Use a ``vertical'' Weyl point annihilation path that passes through the $\Ttori$ $k_z=0$, $k_y=0$ but not $k_x=0$ (Fig.~\ref{fig:familyoftori}(a)-(d)), to obtain a TI with FKMI $\nu_0^{\rm TI},\nu_{i,0}^{\rm TI},\nu_{i,\pi}^{\rm TI}$. Then $\nu_0=\nu_0^{\rm TI}$ and $\nu_{i,\pi}=\nu_{i,\pi}^{\rm TI}$, but $\nu_{i,0}$ requires care. From Fig.~\ref{fig:familyoftori}(b)-(c), only the purple $\Ttori$ persist in the TI and deform to flat ones, whereas the blue ones survive in Fig.~\ref{fig:familyoftori}(d). Thus $\nu_{y/z,0}^{\rm purple}=\nu_{y/z,0}^{\rm TI}=\nu_0^{\rm TI}\nu_{y/z,\pi}^{\rm TI}=\nu_0\nu_{y/z,\pi}$ but $\nu_{x,0}^{\rm blue}=\nu_{x,0}^{\rm TI}=\nu_0^{\rm TI}\nu_{x,\pi}^{\rm TI}=\nu_0\nu_{x,\pi}$, which together with Eq.\ \eqref{blue-purple} gives Eq.\ \eqref{signrelation}. We conclude that $\Twsm$ with $W=\{\pm w, \pm w'\}$ are completely labelled by $(\nu_0;\nu_i;q)\in \ZZ_2^4\oplus\ZZ$ where $\nu_i\equiv\nu_{i,\pi}$. In general, the precise splitting of cases in Eq.\ \eqref{signrelation} also depends on the Weyl point positions $W$ (and a choice of annihilation path for defining $\nu_0$), but it is still true that only three of $\nu_{i,0}, \nu_{i,\pi}$ are independent. Also, if some of the flat $\Ttori$ contain Weyl points, we can replace the ill-defined $\nu_{i,0}, \nu_{i,\pi}$ by the 2D FKMI of slightly curved $\Ttori$ which avoid $W$. 

Formally, a valence band over $X$ without $\mathsf{T}$-symmetry is classified by the \emph{second cohomology group} $H^2(X)$. The analogous group for $\mathsf{T}$-invariant valence bands \cite{dNG,dNG2}, denoted $\mathcal{H}^2(X)$ (see Section \ref{sec:mathequiv}), may be computed (see Section \ref{sec:mathMV}) to be $\mathcal{H}^2(\TT^2)\cong\ZZ_2, \mathcal{H}^2(\TT^3)\cong\ZZ_2^4$ for TI, and $\mathcal{H}^2(\TT^3\setminus W)\cong\ZZ_2^4\oplus\ZZ$ for $\Twsm$, consistent with the above arguments. Abstractly, the dual \emph{homology} groups $\mathcal{H}_1$ do the same job (Section \ref{sec:mathMV}), and we demonstrate this in the next Section, from a physical viewpoint which does not require detailed knowledge of $\mathcal{H}^2, \mathcal{H}_1$.

\section{Semimetal invariants via Weyl point histories}\label{sec:histories}
Define $\nu'_0$ for the $\Twsm$ to be the strong FKMI of the TI that results from annihilating Weyl points \emph{horizontally} (Fig.~\ref{fig:familyoftori}(e)). We expect (from Fig.~\ref{fig:ambiguity}) that $\nu'_0=(-1)^q\nu_0$. Indeed, repeating the above procedures yields $\nu_{x,0}=(-1)^q\nu'_0\nu_{x,\pi}$ and $\nu_{y/z,0}=\nu'_0\nu_{y/z,\pi}$, which is exactly Eq.\ \eqref{signrelation} with $\nu'_0=(-1)^q\nu_0$. 

Next, consider an alternative vertical annihilation path which passes through $k_z=\pi$ (Fig.~\ref{fig:familyoftori}(f)), instead of $k_z=0$ (which the original vertical path passed through). Then it is instead the curved (flat) $\Ttorus$ that corresponds to $\nu_{z,\pi}^{\rm TI}$ ($\nu_{z,0}^{\rm TI}$), so if we wish to end up at the same TI that the original vertical path produced, we need to start from a \emph{different} $\Twsm$ which has its $\nu_{z,0/\pi}$ replaced by $(-1)^q\nu_{z,0/\pi}$; similarly for paths passing through $k_{x/y}=\pi$ instead of $k_{x/y}=0$. 

We conclude that the Weyl point annihilation (creation) path affects the resultant TI ($\Twsm$) topology, as suggested in \cite{HB,OM}. This motivates an alternative classification of TI/$\Twsm$ via Weyl point histories up to $\mathsf{T}$-invariant deformations. 

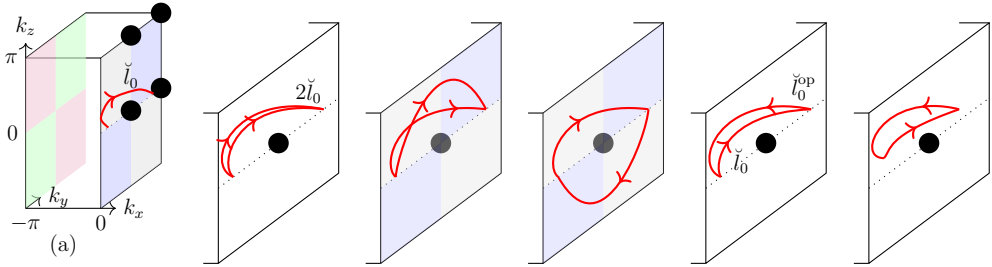
\begin{figure}[bth]
    \centering
\subfigure{
\begin{tikzpicture}[xscale=1, yscale=1,every node/.style={scale=0.7}]

\draw (0,0) -- (1,0) -- (1,2) -- (0,2) -- (0,0);
\draw (0,2) -- (0.8,2.6) -- (1.8,2.6) -- (1,2);
\draw (1,0) -- (1.8,0.6) -- (1.8,2.6);
\draw[dotted] (1,1) -- (1.8,1.6);
\draw[->] (0,2) -- (0,2.2);
\draw[->] (1,0) -- (1.2,0);
\draw[->] (0,0) -- (0.2,0.15);

\node at (0.5,-0.5) {(a)};

\fill[gray!20,opacity=0.6] (1,1) -- (1.4,1.3) -- (1.4,2.3) -- (1,2) -- (1,1);
\fill[blue!20,opacity=0.6] (1.8,1.6) -- (1.4,1.3) -- (1.4,2.3) -- (1.8,2.6) -- (1.8,1.6);
\fill[blue!20,opacity=0.6] (1,1) -- (1.4,1.3) -- (1.4,0.3) -- (1,0) -- (1,1);
\fill[gray!20,opacity=0.6] (1.8,1.6) -- (1.4,1.3) -- (1.4,0.3) -- (1.8,0.6) -- (1.8,1.6);

\fill[purple!20,opacity=0.6] (0,1) -- (0.4,1.3) -- (0.4,2.3) -- (0,2) -- (0,1);
\fill[green!20,opacity=0.6] (0.8,1.6) -- (0.4,1.3) -- (0.4,2.3) -- (0.8,2.6) -- (0.8,1.6);
\fill[green!20,opacity=0.6] (0,1) -- (0.4,1.3) -- (0.4,0.3) -- (0,0) -- (0,1);
\fill[purple!20,opacity=0.6] (0.8,1.6) -- (0.4,1.3) -- (0.4,0.3) -- (0.8,0.6) -- (0.8,1.6);

\node[above] at (0,2.2) {$k_z$};
\node[right] at (1.2,0) {$k_x$};
\node[right] at (0.2,0.15) {$k_y$};

\node[below] at (0,0) {$-\pi$};
\node[left] at (0,1) {$0$};
\node[left] at (0,2) {$\pi$};
\node[below] at (1,0) {$0$};

\node at (1.4,1.3) [circle,fill=black]{} ;
\node at (1.8,1.6) [circle,fill=black]{} ;
\node at (1.4,2.3) [circle,fill=black]{} ;
\node at (1.8,2.6) [circle,fill=black]{} ;

\node at (1.4,1.77) {$\breve{l}_0$};

\draw [red,thick,->-] plot [smooth] coordinates {(1.1,1.075) (1.0,1.2) (1.13,1.45) (1.5,1.58) (1.7,1.525)};

\end{tikzpicture}
}
\subfigure{
\begin{tikzpicture}[scale=1,every node/.style={scale=0.7}]

\draw (-0.2,0) -- (0,0);
\draw (-0.2,2) -- (0,2);
\draw (1.2,3.2) -- (1.6,3.2);
\draw (0,0) -- (0,2) -- (1.6,3.2) -- (1.6,1.2) -- (0,0);
\draw[dotted] (0,1) -- (1.6,2.2);

\node at (0.8,1.6) [circle,fill=black]{} ;

\node at (1.2,2.3) {$2\breve{l}_0$};

\draw[red,thick,->-] (0.2,1.15) .. controls (0,1.3) and (0.2,2.2) .. (1.4,2.05);
\draw[red,thick,->-=0.3] (0.2,1.15) .. controls (-0.2,1.3) and (0.18,2.3) .. (1.4,2.05);

\end{tikzpicture}
}
\hspace{-0.5em}
\subfigure{
\begin{tikzpicture}[scale=1,every node/.style={scale=0.7}]

\draw (-0.2,0) -- (0,0);
\draw (-0.2,2) -- (0,2);
\draw (1.2,3.2) -- (1.6,3.2);
\draw (0,0) -- (0,2) -- (1.6,3.2) -- (1.6,1.2) -- (0,0);
\draw[dotted] (0,1) -- (1.6,2.2);

\node at (0.8,1.6) [circle,fill=black]{} ;

\fill[gray!20,opacity=0.4] (0,1) -- (0.8,1.6) -- (0.8,2.6) -- (0,2) -- (0,1);
\fill[blue!20,opacity=0.4] (1.6,2.2) -- (0.8,1.6) -- (0.8,2.6) -- (1.6,3.2) -- (1.6,2.2);
\fill[blue!20,opacity=0.4] (0,0) -- (0.8,0.6) -- (0.8,1.6) -- (0,1) -- (0,1);
\fill[gray!20,opacity=0.4] (1.6,1.2) -- (0.8,0.6) -- (0.8,1.6) -- (1.6,2.2) -- (1.6,1.2);

\draw[red,thick,->-=0.8] (0.2,1.15) .. controls (0,1.3) and (0.2,2.2) .. (1.4,2.05);
\draw[red,thick,->-] (0.2,1.15) .. controls (0.22,1.5) and (0.8,3.2) .. (1.4,2.05);

\end{tikzpicture}
}
\hspace{-0.5em}
\subfigure{
\begin{tikzpicture}[scale=1,every node/.style={scale=0.7}]

\draw (-0.2,0) -- (0,0);
\draw (-0.2,2) -- (0,2);
\draw (1.2,3.2) -- (1.6,3.2);
\draw (0,0) -- (0,2) -- (1.6,3.2) -- (1.6,1.2) -- (0,0);
\draw[dotted] (0,1) -- (1.6,2.2);

\node at (0.8,1.6) [circle,fill=black]{} ;

\fill[gray!20,opacity=0.4] (0,1) -- (0.8,1.6) -- (0.8,2.6) -- (0,2) -- (0,1);
\fill[blue!20,opacity=0.4] (1.6,2.2) -- (0.8,1.6) -- (0.8,2.6) -- (1.6,3.2) -- (1.6,2.2);
\fill[blue!20,opacity=0.4] (0,0) -- (0.8,0.6) -- (0.8,1.6) -- (0,1) -- (0,1);
\fill[gray!20,opacity=0.4] (1.6,1.2) -- (0.8,0.6) -- (0.8,1.6) -- (1.6,2.2) -- (1.6,1.2);

\draw[red,thick,->-] (0.2,1.15) .. controls (0,1.3) and (0.2,2.2) .. (1.4,2.05);
\draw[red,thick,->-] (1.4,2.05) .. controls (1.38,1.7) and (0.8,0.1) .. (0.2,1.15);

\end{tikzpicture}
}
\hspace{-0.5em}
\subfigure{
\begin{tikzpicture}[scale=1,every node/.style={scale=0.7}]

\draw (-0.2,0) -- (0,0);
\draw (-0.2,2) -- (0,2);
\draw (1.2,3.2) -- (1.6,3.2);
\draw (0,0) -- (0,2) -- (1.6,3.2) -- (1.6,1.2) -- (0,0);
\draw[dotted] (0,1) -- (1.6,2.2);

\node at (0.8,1.6) [circle,fill=black]{} ;

\node at (1.3,2.35) {$\breve{l}_0^{\rm op}$};
\node at (0.47,1.37) {$\breve{l}_0$};

\draw[red,thick,->-=0.3] (0.2,1.15) .. controls (0.1,1.3) and (0.2,2) .. (1.4,2.05);
\draw[red,thick,->-=0.3] (1.4,2.05) .. controls (0.18,2.3) and (-0.2,1.3) .. (0.2,1.15);

\end{tikzpicture}
}
\hspace{-0.5em}
\subfigure{
\begin{tikzpicture}[scale=1,every node/.style={scale=0.7}]

\draw (-0.2,0) -- (0,0);
\draw (-0.2,2) -- (0,2);
\draw (1.2,3.2) -- (1.6,3.2);
\draw (0,0) -- (0,2) -- (1.6,3.2) -- (1.6,1.2) -- (0,0);
\draw[dotted] (0,1) -- (1.6,2.2);

\node at (0.8,1.6) [circle,fill=black]{} ;

\draw[red,thick,->-=0.5] (0.2,1.4) .. controls (0.3,1.5) and (0.2,1.7) .. (1.2,2.05);
\draw[red,thick,->-=0.3] (1.2,2.05) .. controls (0.18,2.3) and (-0.2,1.4) .. (0.2,1.4);

\end{tikzpicture}
}

\caption{(a) The quotient $\breve{\TT}^3$ of the BZ under $\tau$ is the 3D EBZ with further identifications $(k_y,k_z)\sim (-k_y,-k_z)$ at each EBZ boundary face $k_x=0,-\pi$. Since its end points are identified, the red arc is a closed loop $\breve{l}_0$ in $\breve{\TT}^3$, which is ``half'' of the equator $l_0$ surrounding $\vect{0}$. Starting from \emph{two} such loops, deform one $\breve{l}_0$ into a loop on the $k_x=0$ face, identify it with its inverted self, then deform it back to $\breve{l}_0^{\rm op}$. Thus $2\breve{l}_0\sim\breve{l}_0+\breve{l}^{\rm op}_0$, which is contractible in $\breve{\TT}^3$.}\label{fig:torsionloops}

\end{figure}

Closed arcs/paths (i.e.\ loops) $l$ in a space $X$ are classified up to deformation by the first \emph{homology group} $H_1(X)$\footnote{Strictly speaking, closed paths are classified by the first \emph{homotopy group}, but if we allow reordering of concatenated paths (\emph{abelianization}), we recover $H_1(X)$. Physically, the loops projected to the surface BZ correspond to the Fermi \emph{locus} and the ordering does not matter.}; e.g.\ there are three independent non-contractible loops in $\TT^3$, so $H_1(\TT^3)\cong\ZZ^3$. For WSM, arcs connecting Weyl points are allowed (``$W$-relative loops''\footnote{Arcs connecting Weyl points can be thought of as loops upon identifying points in $W$.}), so the \emph{relative homology} group $H_1(\TT^3,W)$ is required. For TI/$\Twsm$, \emph{$\tau$-equivariant} homology groups (Section \ref{sec:mathequiv}), denoted $\mathcal{H}_1(\TT^3), \mathcal{H}_1(\TT^3,W)$, classify $\Tinv$ loops. Equivalently, we can classify loops $\breve{l}$ in the \emph{quotient space} $\breve{\TT}^3$ of the BZ under $\tau$. This quotient space is nothing but the 3D EBZ with additional $k\sim \tau(k)$ identifications at its two boundary faces (Fig.\ \ref{fig:torsionloops}(a)). A loop $\breve{l}$ in $\breve{\TT}^3$ ``doubles'' into a $\Tinv$ loop $l=\breve{l}\cup\tau(\breve{l})$ in $\TT^3$. 

Each pair $(k,\tau(k))$ in the BZ is ``halved'' into a single point in $\breve{\TT}^3$, with the important caveat that ``halving a TRIM'' technically requires the \emph{homotopy quotient} rather than $\breve{\TT}^3$. For our purposes, it suffices to look at loops in $\breve{\TT}^3$ (or $\Tinv$ loops in $\TT^3$) which \emph{avoid} the TRIM. We will see that $\mathcal{H}_1(\TT^3)\cong\ZZ_2^4$ and $\mathcal{H}_1(\TT^3,W)\cong\ZZ_2^4\oplus\ZZ$, exactly mirroring the classification of TI and $\Twsm$ (by $\mathcal{H}^2$). This apparent coincidence is due to a deep result called \emph{Poincar\'{e} duality} (PD) \cite{BottTu,Hatcher} (Section \ref{sec:mathpoincare}), see \cite{Gross} for PD applied to electromagnetic theory. In the $\mathsf{T}$-broken case, PD identifies $H^2(\TT^3)\cong H_1(\TT^3)\cong\ZZ^3$. For the $\mathsf{T}$-symmetric case, PD identifies $\mathcal{H}^2(\TT^3)$ and $\mathcal{H}^2(\TT^3\setminus W)$ with $\mathcal{H}_1(\TT^3)$ and $\mathcal{H}_1(\TT^3,W)$ \cite{GT}.

\begin{figure}[bth]
    \centering
\subfigure{
\begin{tikzpicture}[scale=1, every node/.style={scale=0.7}]
\draw[->](-0.4,1.7) -- (-0.4,1.9);
\draw[->](1.6,-0.3) -- (1.8,-0.3);
\draw[->](-0.4,-0.3) -- (-0.2,-0.15);

\draw (-0.4,-0.3) -- (-0.4,1.7) -- (1.6,1.7) -- (1.6,-0.3) -- (-0.4,-0.3);
\draw (-0.4,1.7)  -- (0.4,2.3) -- (2.4,2.3) -- (2.4,0.3) -- (1.6,-0.3);
\draw (1.6,1.7) -- (2.4,2.3);

\node at (-0.4,2.3) {(a)};

\node [left] at (-0.4,1.9) {$k_z$} ;
\node [below] at (1.9,-0.3) {$k_x$} ;
\node [right] at (-0.2,-0.15) {$k_y$};
\node [left] at (-0.4,1.6) {$\pi$} ;
\node [left] at (-0.4,0.7) {$0$} ;
\node [below] at (0.6,-0.3) {$0$} ;
\node [below] at (1.6,-0.3) {$\pi$} ;
\node at (1,1) [circle,fill=black]{} ;
\node at (2,1) [circle,fill=black]{} ;
\node at (1.4,1.3) [circle,fill=black]{} ;
\node at (2.4,1.3) [circle,fill=black]{} ;

\draw [red,thick,->-]  (1.2,1.65) -- (0.4,1.05);
\draw [red,thick,->-] (0.8,0.35) -- (1.6,0.95);
\draw[dashed,->=0.9,domain=-0.5:0.5] plot (0.5*\x+0.6, {0.7+(0.25-\x*\x)^(0.5) });
\draw[dashed,<-=0.9,domain=-0.5:0.5] plot (0.5*\x+0.6, {0.7-(0.25-\x*\x)^(0.5) });
\draw[dashed,->=0.9,domain=-0.5:0.5] plot (0.5*\x+1.4, {1.3+(0.25-\x*\x)^(0.5) });
\draw[dashed,<-=0.9,domain=-0.5:0.5] plot (0.5*\x+1.4, {1.3-(0.25-\x*\x)^(0.5) });

\end{tikzpicture}
}
\hspace{-0.5em}
\subfigure{
\begin{tikzpicture}[scale=1.7, every node/.style={scale=0.7}]


\node at (1,1.7) {(b)};
\node at (1,0) {};

\node at (1,1) [circle,fill=black]{} ;

\draw[thick,dashed,->=0.5,domain=-0.5:0.5] plot (1+\x, {1+(0.25-\x*\x)^(0.5) });
\draw[thick,dashed,<-=0.5,domain=-0.5:0.5] plot (1+\x, {1-(0.25-\x*\x)^(0.5) });

\draw[red,thick,dotted] (0.6,1.3) to [bend left=45] (1.4,0.7);
\draw[red,thick,->-=0.3] (1.4,0.7) to [bend left=45] (0.6,1.3);

\draw[blue,dotted] (1.4,1.3) to [bend left=45] (0.6,0.7);
\draw[blue,->-=0.7] (0.6,0.7) to [bend left=45] (1.4,1.3);


\end{tikzpicture}
}
\hspace{-0.5em}
\subfigure{
\begin{tikzpicture}[scale=1, every node/.style={scale=0.7}]
\draw (0,0) -- (2,0) -- (2,2) -- (0,2) -- (0,0);
\draw [->] (0,2) -- (0,2.2);
\draw [->] (2,0) -- (2.2,0);
\draw [->] (0,0) -- (0.2,0.15);

\node [above] at (0.2,0.13) {$k_y$};
\node [right] at (0,2.2) {$k_z$} ;
\node [below] at (2.15,0) {$k_x$} ;
\node [below] at (1.85,0) {$\pi$} ;
\node [left] at (0,2) {$\pi$} ;
\node[left] at (0,1) {$0$};
\node [below] at (1,0) {$0$};

\node at (1,1) [circle,fill=black]{} ;
\node at (2,1) [circle,fill=black]{} ;
\node at (2,2) [circle,fill=black]{} ;
\node at (1,2) [circle,fill=black]{} ;

\draw[red,thick,->-,domain=0.5:-0.5] plot (1+\x, {(0.25-\x*\x)^(0.5) });
\draw[red,thick,->-,domain=-0.5:0.5] plot (1+\x, {2-(0.25-\x*\x)^(0.5) });

\draw [magenta,->-=0.6] plot [smooth] coordinates {(0.5,2) (0.6,1.3) (1,1.4) (1.4,1.3) (1.5,2)};
\draw [magenta,->-=0.6] plot [smooth] coordinates {(1.5,0) (1.4,0.7) (1,0.6) (0.6,0.7) (0.5,0)};

\draw [blue,->-=0.45] (1.5,0) -- (1.5,2);
\draw [blue,->-=0.65] (0.5,2) -- (0.5,0);
\draw [blue,->-,domain=-0.5:0.5] plot (1+\x, {1+0.5*(0.25-\x*\x)^(0.5) });
\draw [blue,domain=-0.5:0.5] plot (1+\x, {1-0.5*(0.25-\x*\x)^(0.5) });

\node at (1,2.3) {(c)};







\end{tikzpicture}
}
\hspace{-0.5em}
\subfigure{
\begin{tikzpicture}[scale=0.8, every node/.style={scale=0.7}]
\draw[domain=-1:1] plot (\x, {2.5+0.25*(1-\x*\x)^(0.5) });
\draw[domain=-1:1] plot (\x, {2.5-0.25*(1-\x*\x)^(0.5) });
\draw[dotted,domain=-1:1] plot (\x, {0.25*(1-\x*\x)^(0.5) });
\draw[domain=-1:1] plot (\x, {-0.25*(1-\x*\x)^(0.5) });
\draw (-1,0) -- (1,2.5);
\draw (1,0) -- (-1,2.5);
\draw[red,thick,->=0.3] (0,2.25)--(0.3,2.25);
\draw[red,thick,->=0.3] (0,2.75)--(-0.3,2.75);
\draw[red,thick,->=0.3] (0.7,2.3)--(0.9,2.35);
\draw[red,thick,->=0.3] (0.7,2.7)--(0.5,2.75);
\draw[red,thick,->=0.3] (-0.7,2.3)--(-0.5,2.25);
\draw[red,thick,->=0.3] (-0.7,2.7)--(-0.9,2.65);

\draw[red,dotted,thick,->=0.3] (0,0.25)--(0.3,0.25);
\draw[red,thick,->=0.3] (0,-0.25)--(-0.3,-0.25);
\draw[red,dotted,thick,->=0.3] (0.7,-0.2)--(0.5,-0.25);
\draw[red,dotted,thick,->=0.3] (0.7,0.2)--(0.9,0.15);
\draw[red,thick,->=0.3] (-0.7,0.2)--(-0.5,0.25);
\draw[red,thick,->=0.3] (-0.7,-0.2)--(-0.9,-0.15);

\node at (0,3.3) {(d)};
\node at (0,-0.5) {};

\draw [->=0.3] (1.3,0) -- (1.3,3);
\node[left] at (1.3,3.1) {$E$};

\draw[brown,thick,dashed] (-1.3,0.9) -- (0.9,0.9) -- (1.3,1.6) -- (-0.9,1.6) -- (-1.3,0.9);
\draw[brown,thick,dashed] (-1.3,2.15) -- (0.9,2.15) -- (1.3,2.85) -- (-0.9,2.85) -- (-1.3,2.15);
\draw[brown,thick,dashed] (-1.3,-0.35) -- (0.9,-0.35) -- (1.3,0.35) -- (-0.9,0.35) -- (-1.3,-0.35);

\end{tikzpicture}
}

\subfigure{
\begin{tikzpicture}[scale=1, every node/.style={scale=0.7}]
\draw (0,0) -- (2,0) -- (2,2) -- (0,2) -- (0,0);
\draw [->] (0,2) -- (0,2.2);
\draw [->] (2,0) -- (2.15,0);
\draw [->] (0,0) -- (0.2,0.15);

\node [right] at (0,2.25) {$k_z$} ;
\node [above] at (2.15,0) {$k_x$} ;
\node [right] at (0.2,0.15) {$k_y$} ;

\node at (1,1) [circle,fill=black]{} ;
\node at (2,1) [circle,fill=black]{} ;
\node at (2,2) [circle,fill=black]{} ;
\node at (1,2) [circle,fill=black]{} ;
\node at (0.6,1.5) {$\boldsymbol{\oplus}$};
\node at (0.4,0.7) {$\boldsymbol{\ominus}$};
\node at (1.4,0.5) {$\boldsymbol{\oplus}$};
\node at (1.6,1.3) {$\boldsymbol{\ominus}$};
\node[above] at (0.6,1.5) {$w$};
\node[right] at (0.42,0.7) {$w'$};
\node[below] at (1.4,0.5) {$-w$};
\node[above] at (1.6,1.3) {$-w'$};

\draw[red,->-] (0.4,0.7) to  [bend right] (0.6,1.5);
\draw[red,->-] (1.6,1.3) to  [bend right] (1.4,0.5);

\end{tikzpicture}
}
\hspace{-1em}
\subfigure{
\begin{tikzpicture}[scale=1, every node/.style={scale=0.7}]
\draw (0,0) -- (2,0) -- (2,2) -- (0,2) -- (0,0);
\draw [->] (0,2) -- (0,2.2);
\draw [->] (2,0) -- (2.15,0);


\node at (1,1) [circle,fill=black]{} ;
\node at (2,1) [circle,fill=black]{} ;
\node at (2,2) [circle,fill=black]{} ;
\node at (1,2) [circle,fill=black]{} ;
\node at (0.6,1.5) {$\boldsymbol{\oplus}$};
\node at (0.4,0.7) {$\boldsymbol{\ominus}$};
\node at (1.4,0.5) {$\boldsymbol{\oplus}$};
\node at (1.6,1.3) {$\boldsymbol{\ominus}$};

\draw [red,->-] plot [smooth] coordinates {(0.4,0.7) (0.9,0.8) (0.7,1) (0.9,1.4) (0.6,1.5)};
\draw [red,->-] plot [smooth] coordinates {(1.6,1.3) (1.1,1.2) (1.3,1) (1.1,0.6) (1.4,0.5)};

\node at (2,2.4) {(e)};

\end{tikzpicture}
}
\hspace{-1em}
\subfigure{
\begin{tikzpicture}[scale=1, every node/.style={scale=0.7}]
\draw (0,0) -- (2,0) -- (2,2) -- (0,2) -- (0,0);
\draw [->] (0,2) -- (0,2.2);
\draw [->] (2,0) -- (2.15,0);


\node at (1,1) [circle,fill=black]{} ;
\node at (2,1) [circle,fill=black]{} ;
\node at (2,2) [circle,fill=black]{} ;
\node at (1,2) [circle,fill=black]{} ;
\node at (0.6,1.5) {$\boldsymbol{\oplus}$};
\node at (0.4,0.7) {$\boldsymbol{\ominus}$};
\node at (1.4,0.5) {$\boldsymbol{\oplus}$};
\node at (1.6,1.3) {$\boldsymbol{\ominus}$};

\draw [red,->-] (0.4,0.7) to [bend left] (1.4,0.5);
\draw [red,->-] (1.6,1.3) to [bend left] (0.6,1.5);
\draw[red,->-=0.2,domain=0.23:-0.23] plot (1+\x, {1-(0.23*0.23-\x*\x)^(0.5) });
\draw[red,->-=0.2,domain=-0.23:0.23] plot (1+\x, {1+(0.23*0.23-\x*\x)^(0.5) });

\end{tikzpicture}
}
\hspace{-1em}
\subfigure{
\begin{tikzpicture}[scale=1, every node/.style={scale=0.7}]
\draw (0,0) -- (2,0) -- (2,2) -- (0,2) -- (0,0);
\draw [->] (0,2) -- (0,2.2);
\draw [->] (2,0) -- (2.15,0);


\node at (1,1) [circle,fill=black]{} ;
\node at (2,1) [circle,fill=black]{} ;
\node at (2,2) [circle,fill=black]{} ;
\node at (1,2) [circle,fill=black]{} ;
\node at (0.6,1.5) {$\boldsymbol{\oplus}$};
\node at (0.4,0.7) {$\boldsymbol{\ominus}$};
\node at (1.4,0.5) {$\boldsymbol{\oplus}$};
\node at (1.6,1.3) {$\boldsymbol{\ominus}$};

\draw[red,->-] (0.4,0.7) to  [bend left=10] (1.4,0.5);
\draw[red,->-] (1.6,1.3) to  [bend left=10] (0.6,1.5);
\draw[red,->-,domain=-0.2:0.2] plot (1+\x, {1+(0.04-\x*\x)^(0.5) });
\draw[red,->-,domain=0.2:-0.2] plot (1+\x, {1-(0.04-\x*\x)^(0.5) });

\end{tikzpicture}
}
\caption{(a) A $\Tinv$ pair of loops (red, dual to $\nu_y$) rotates about the $k_y$-axis onto its oppositely-oriented self. (b) Red equator encircling TRIM at $\vect{0}$, dual to $\nu_0$, rotates onto any other equator (blue). (c) Red equator surrounding $(0,0,\pi)$ TRIM deforms into sum of blue equator around $\vect{0}$ and blue loops along $k_z$. (d) Helical Dirac cone with spin texture (red). (e) Vertical Weyl point connection deforms into horizontal connection plus a circle dual to $\nu_0$, numerically reproduced in Fig.\ \ref{fig:ABstack}(d).}
\label{fig:dualFKM}

\end{figure}
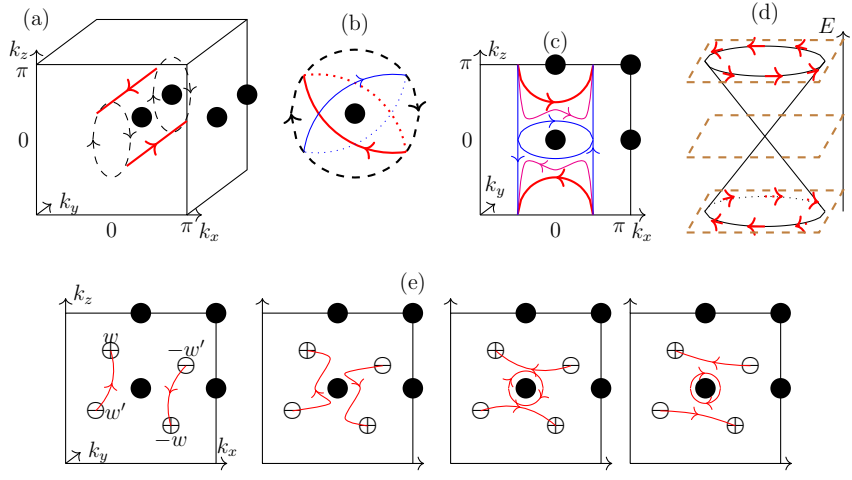

For each $i=x,y,z$, let $l_i$ be a $\Tinv$ pair of loops winding around $k_i$ (one loop in the pair is the $\tau$-image of the other), and consider its deformations through $\Tinv$ loops. It cannot be contracted to a point without hitting TRIMs, but does rotate onto its oppositely-oriented self $l_i^{\rm op}$ (Fig.\ \ref{fig:dualFKM}(a)). Thus \emph{two} copies of $l_i$ deform into $l_i+l_i^{\rm op}$, which \emph{together} is $\mathsf{T}$-stably contractible, so $l_i$ generates $\ZZ_2\subset\mathcal{H}_1(\TT^3)$; this $\ZZ_2$ is dual to the weak FKMI $\nu_i$. Next, a $\Tinv$ equator $l_0$ of a 2-sphere surrounding the TRIM at $\vect{0}$ cannot be contracted \emph{$\mathsf{T}$-stably} without hitting the TRIM. All equators are equivalent since one can be rotated onto another (Fig.\ \ref{fig:dualFKM}(b)); in particular, $l_0$ is equivalent $l_0^{\rm op}$, so $l_0$ also generates a $\ZZ_2\subset\mathcal{H}_1(\TT^3)$, dual to the strong FKMI $\nu_0$. Projected onto the surface BZ, $l_0$ gives the Fermi surface of a surface Dirac cone pinned to $(0,0)$ while $l_i$ gives that of weak TI exactly as in \cite{FKM,Hasan}. An equator encircling any other TRIM is not another independent loop, but is deformable to some combination of $l_0$ and $l_i$ (Fig.\ \ref{fig:dualFKM}(c)), thus $\mathcal{H}_1(\TT^3)\cong\ZZ_2^4$. For $\Twsm$ with $W=\{\pm w, \pm w'\}$, a $\Tinv$ pair of vertical arcs joining $\pm w'$ to $\pm w$ generates $\ZZ$ in $\mathcal{H}_1(\TT^3,W)\cong\ZZ_2^4\oplus\ZZ$, dual to $q$, while the horizontal arc-pair joining $\pm w'$ to $\mp w$ is a combination of the vertical one and $l_0$ (Fig.\  \ref{fig:dualFKM}(e)). The Weyl point connections then prefigure the surface Fermi arcs by a simple projection onto the surface BZ.

We can also see why $l_0, l_i$ each generates $\ZZ_2$, by considering their respective ``halves'' $\breve{l}_0,\breve{l}_i$ in $\breve{\TT}^3$. For example, Fig.\ \ref{fig:torsionloops} shows how a loop going around $\breve{l}_0$ \emph{twice} is contractible, see Fig.\ \ref{fig:torsionloops2} for the $\breve{l}_i$ case. This is just like $\text{SO}(3)$, the quotient of $\text{SU}(2)$ by $\ZZ_2$, having a loop which only becomes contractible after looping twice\footnote{This analogy is actually precise: a TRIM is properly replaced in the homotopy quotient E$\ZZ_2\times_{\ZZ_2}\TT^3$ by the classifying space B$\ZZ_2\cong\RR\PP^\infty$ (the infinite real projective space). Due to the $\RR\PP^3\cong\text{SO}(3)$ part, there is a non-contractible loop in the homotopy quotient which becomes contractible after looping twice.}.

In this arc-representation of TI/$\Twsm$, the FKMI on any $\Ttorus$ is simply the number of intersection-pairs (mod 2) between the arcs and the $\Ttorus$, e.g.\ the red loops in Fig.\ \ref{fig:dualFKM}(a) represent a weak TI with $(\nu_0;\nu_i)=(+;+,-,+)$, while the equator around $(0,0,\pi)$ represents a strong TI with $(\nu_0;\nu_
i)=(-;+,+,-)$ (Fig.\ \ref{fig:dualFKM}(c)). Intersection numbers are another formulation of PD, were used implicitly in characterising the Fermi surface on the surface of a TI by intersection properties with certain $\tau$-stable lines in the surface BZ \cite{FK,FKM,Hasan}, and were applied explicitly to  semimetals in \cite{MT1}.

\section{Bulk-boundary map}\label{sec:bbc}
In the $\mathsf{T}$-broken case, the bulk-boundary correspondence involves integrating a bulk quantity (e.g.\ 2D Berry curvature along, say, $k_y$) transversally to the boundary, to obtain a boundary topological invariant (e.g.\ 1D winding number), c.f.\ dimensional reduction \cite{QHZ}, charge pumping \cite{Thouless}, Zak phase computation \cite{Kim}, spectral flow/Toeplitz index \cite{Hatsugai,GP,ASV,KSB}, and Section \ref{sec:mathbbc}. With $\mathsf{T}$-symmetry, we need to ``integrate out $k_y$ $\mathsf{T}$-invariantly'', c.f.\ the spin pump \cite{FK} generalising \cite{Thouless}). 

Integration of a differential form along $k_y$ transforms under PD into the map $p:\TT^3\rightarrow\TT^2$ projecting out $k_y$ \cite{BottTu}, see Section \ref{sec:mathpoincare}. This indicates why the Fermi surface in the surface BZ can be found by dualising a $\Twsm$ into arcs and then projecting onto the surface BZ,
\begin{equation*}
\mathcal{P}:\underbrace{\mathcal{H}^2(\TT^3\setminus W)}_{\rm Bulk\, invariants}\underset{\rm PD}{\xrightarrow{\sim}}\underbrace{\mathcal{H}_1(\TT^3,W)}_{\rm arc\,representation}\xrightarrow{p}\underbrace{\mathcal{H}_1(\TT^2,p(W))}_{\rm boundary\, invariants}.
\end{equation*}
Note that the codomain $\mathcal{H}_1(\TT^2,p(W))$ classifies the possible $\Tinv$ Fermi surfaces on the surface BZ, i.e.\ closed loops and/or open Fermi arcs connecting projected Weyl points.

The $\Twsm$ generically has four projected Weyl points, then $\mathcal{H}_1(\TT^2,p(W))\cong\ZZ_2\oplus\ZZ_2^2\oplus\ZZ$: the first factor is $\mathcal{P}(\nu_0)$ (loop around $(k_x,k_z)=(0,0)$), the middle factor is $\mathcal{P}(\nu_x), \mathcal{P}(\nu_z)$ (loop-pairs wrapping $k_x,k_z$), the last factor is $\mathcal{P}(q)$ (Fermi arc-pair from $p(\pm w')$ to $p(\pm w)$), while $\nu_y$ does not contribute since it corresponds to a weak TI built from layers of 2D TI parallel to the surface.

Recall that the \emph{helical} Dirac cone of a strong TI has spin-momentum locking in opposite senses for the two halves of the cone \cite{Hsieh, Hasan} (Fig.\ \ref{fig:dualFKM}(d)). As the Fermi energy is varied slightly, the Fermi surface (a loop oriented along the spin) undergoes a ``Lifshitz transition'' first to a Dirac point, then to an oppositely-oriented loop. All these Fermi surfaces arise from the same bulk Hamiltonian, and their interpolation as mediated by the Dirac cone realises their topological equivalence. The Dirac point (as a degenerate loop) is \emph{essential} for the Fermi loop orientation to change continuously so that such a loop has a $\ZZ_2$ index.

Under a continuous change of the bulk Hamiltonian, neither the $\Twsm$ indices nor the dual arcs' topology changes. It is then easy, for more complicated Weyl point configurations, to recognise which Fermi arc-Dirac cone configurations are continuous ``rewirings'' of one another (by deforming the $\Twsm$ Hamiltonian whilst keeping the Weyl points fixed) \cite{Dwivedi}, and which ones require a genuine topological phase transition \cite{MT1} of $\Twsm$.

\subsection{Interface of semimetals} 
\begin{figure}[bth]
\centering
\includegraphics[width=0.55\linewidth]{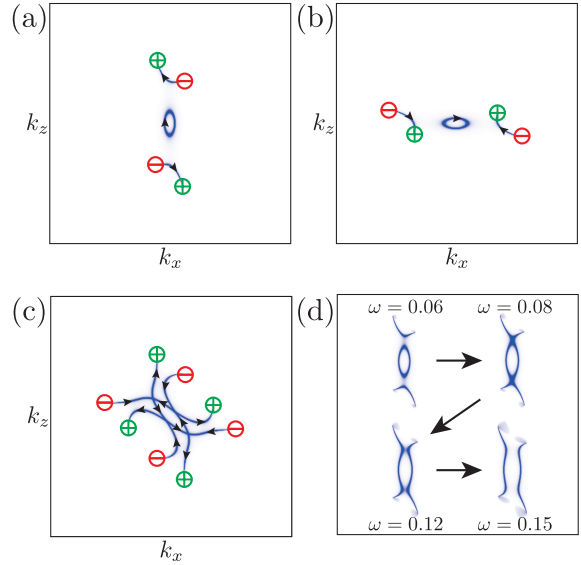}
\caption{\label{fig:ABstack}(a) Surface Fermi surface of $\Twsm$ A corresponding to the right most figure in Fig.~3(e) with two Fermi arcs and a single Dirac cone. (b) Fermi surface on opposite surface of $\Twsm$ B, obtained from A by reflecting. (c) Interface Fermi surface of the system obtained by stacking A and B. (d) Fermi surfaces of A at various Fermi energies $\omega$, reproducing the deformation sequence of Fig.~3(e).}

\end{figure}

Consider two distinct $\Twsm$ A and B, with respective indices $(+;+,+,+;1)$ and $(-;+,+,+;1)$. At their interface (transverse to $y$), we expect the Fermi surface to be the \emph{difference} $\mathcal{P}(+;+,+,+;1)-\mathcal{P}(-;+,+,+;1)=(-;+,+;0)$, i.e., a single Dirac cone. Using Hamiltonians from \cite{Lau} to realise A (resp.\ B) with Fermi surfaces comprising a Dirac cone and horizontally (resp.\ vertically) connected Fermi arcs respectively, as in Fig.~\ref{fig:ABstack}(a)-(b), we numerically analyse, using Green's function methods \cite{Sancho}, the interface-localised Fermi surface of the system with A stacked on top of B in the $y$-direction, obtaining Fig.~\ref{fig:ABstack}(c). Instead of the Dirac cone from A annihilating that from B, we see that there is still a loop around $(k_x,k_z)=(0,0)$ corresponding to a single Dirac cone, while the remaining ``rewired'' Fermi arcs can be thought of as being contractible if the projected Weyl points of A had lined up exactly with those of B. This verifies our topological classification via $\mathcal{H}_1$, and demonstrates that the correspondence between the mod 2 number of Dirac cones and the strong invariant becomes more subtle in the presence of Fermi arcs.

\section{Mathematical addendum}\label{sec:math}
\subsection{Homology, Fermi surfaces, and arcs}\label{sec:mathhomology}
The $n$-th homology group $H_n(X)$ of a space $X$ is a topological invariant which detects ``$n$-dimensional holes'' in $X$ \cite{Hatcher, BottTu}. For example, $\TT^2$ has two ``1-dimensional holes'' since the loops wrapping $k_i, i=x,y$ cannot be contracted to a point within the BZ. This topological property is encoded by an abelian group $H_1(\TT^2)\cong\ZZ^2$, and is relevant e.g.\ for classifying Fermi surfaces in 2D, which are closed 1-submanifolds that may wrap around such ``holes''. Strictly speaking, non-contractible loops in $X$ are classified by the first \emph{homotopy group} $\pi_1(X)$, which is not generally abelian, but if we forget the ordering of loops (abelianisation), we do recover $H_1(X)$.

$H_1(X)$ is more precisely defined by considering continuous maps $l$ from a 1-simplex $[0,1]$ into $X$ (i.e.\ arcs that ``sample'' the topology of $X$). The \emph{boundary} $\partial l$ of $l$ is the formal difference $l(1)-l(0)$. A \emph{1-chain} is a formal sum $L=\sum_i m_i\cdot l_i, m_i\in\ZZ$, and is called a \emph{1-cycle} if its boundary $\partial L\equiv \sum_i m_i\cdot\partial l_i$ vanishes (e.g. if $L$ is actually a single map from a \emph{circle} into $X$). $H_1(X)$ is then the 1-cycles modulo those which are obtainable from (some formal sum of) maps $s_j:2$-simplex$\rightarrow X$ by restriction to the boundary 1-simplices of $s_j$ (such 1-cycles are called \emph{1-boundaries}). In $H_1(X)$, the inverse of $l$ can be represented by the same map but going around the circle in the opposite sense, denoted $\l^{\rm op}$. The higher-degree groups $H_n(X)$ are defined in an analogous way using maps from $n$-simplices into $X$ (\emph{$n$-chains}). If the coefficients $m_i$ come from some other abelian group $G$, we get the \emph{homology groups with $G$-coefficients} $H_n(X;G)$. By default, $H_n(X)\equiv H_n(X;\ZZ)$.

\subsection{Cohomology, Berry curvature, and valence bands}\label{sec:mathcohomology}
Cohomology groups with $G$-coefficients, denoted $H^n(X;G)$, are generally defined dually to homology, i.e.\ using homomorphisms from $n$-chains into $G$ (called \emph{$n$-cochains}), and there are \emph{coboundary} maps $d$, \emph{cocycles} etc. These groups are important for studying topological insulators/semimetals because they classify various types of vector bundles (or valence bands with various symmetries). For example, bundles of one-dimensional complex (resp.\ real) vector spaces over $X$ are classified by $H^2(X;\ZZ)$, (resp.\ $H^1(X,\ZZ_2)$), and non-trivial 2D Chern insulators are possible due to $H^2(\TT^2;\ZZ)\cong\ZZ$. For smooth manifolds $X$, the \emph{de Rham cohomology groups} $H_{\rm de\, Rham}^n(X;\RR)$ provide a more familiar model for the case $G=\RR$; they are given by the closed differential $n$-forms ($d\omega=0$) modulo the exact ones ($\omega=d\eta$ for some $\eta$). The inclusion of coefficients $\ZZ\rightarrow\RR$ induces a map $H^n(X;\ZZ)\rightarrow H^n(X;\RR)$ so that we may study some (but not all) aspects of $H^n(X;\ZZ)$ using differential forms that integrate to integers on closed submanifolds. A typical example is the representation of the Chern class (a priori an element of $H^2(X,\ZZ)$) of a valence band by its Berry curvature 2-form $\mathcal{F}$ (which gives an element of $H^2(X,\RR)$).

\subsection{(Co)homology with $\mathsf{T}$-symmetry}\label{sec:mathequiv}
Fermionic $\mathsf{T}$-symmetry ($\mathsf{T}^2=-1$) furnishes valence bands with a ``Quaternionic'' structure with respect to the momentum-reversing map $\tau$, i.e.\ it relates $|\psi(k)\rangle$ to $|\psi(\tau(k))\rangle$ antiunitarily (Kramers pair), and squares to $-1$. To classify such ``Quaternionic'' vector bundles, we need to use modified cohomology groups $\mathcal{H}^2(X)$ \cite{dNG, dNG2}, referred to in the main text, which are defined for every space $X$ with $\ZZ_2$-action $\tau$ and fixed point set $F$. An example is $X$ a Brillouin torus, $\tau:k\mapsto -k$, and $F$ the TRIM set. More precisely, $\mathcal{H}^n(X)\equiv H^n_{\ZZ_2}(X,F;\wt{\ZZ})$, which is a $\ZZ_2$-\emph{equivariant} cohomology group \emph{relative} to $F$ with \emph{local coefficients} $\wt{\ZZ}$. The meaning of $\wt{\ZZ}$ is that $\ZZ_2$ also acts on the coefficients $\ZZ$ by $m\mapsto -m$, and ``equivariant'' means that cochains are required to satisfy a certain $\ZZ_2$-symmetry condition with respect to $\tau$. These modifications keep track of the Kramers pair condition so that a ``Quaternionic'' vector bundle has a characteristic class in $\mathcal{H}^2(X)$ \cite{dNG,dNG2}. Up to some mild technical assumptions, $\mathcal{H}^2(X)$ classifies all possible $\mathsf{T}$-invariant valence bands over $X$ if dim($X)\leq3$ (with given even number of bands) \cite{dNG, dNG2}. 

The modified homology group $\mathcal{H}_n(X)$, referred to in the main text, is significantly simpler --- it is the \emph{equivariant} homology group $H_n^{\ZZ_2}(X;\ZZ)$ with ordinary integer coefficients. Intuitively, instead of probing $X$ by mapping loops, surfaces, simplices etc.\ into $X$ as in ordinary homology, we need to start with spaces $Z$ that themselves have a $\ZZ_2$ action, and ``probe equivariantly''. For example, the 1-chains in $\mathcal{H}_1(X)$ are, roughly speaking, maps $l$ from 1D spaces $Z$ with free involution $\tau_{\rm free}$ (i.e.\ having no fixed points) into $X$ which satisfy $\tau(l(z))=l(\tau_{\rm free}(z))$. Thus each pair of points $z,\tau_{\rm free}(z)$ is mapped onto a conjugate pair $x,\tau(x)$ (which may coincide), and $l$ traces out a $\Tinv$ 1D subspace of $X$. For $\Tinv$ subspaces $Y\subset X$ (e.g. the Weyl point set $W$), the \emph{relative} homology group $\mathcal{H}_1(X,Y)$ allows for $\Tinv$ \emph{open} arcs that start and end in $Y$. In this way, $\mathcal{H}_1(X,W)$ is relevant for classifying $\Tinv$ Fermi arcs.

If $X$ has no $\tau$-fixed points (which is atypical), the equivariant homology $H^{\ZZ_2}_n(X)\equiv \mathcal{H}_n(X)$ reduces to the \emph{ordinary} homology of the na\"{i}ve quotient space $\breve{X}$ obtained from $X$ by identifying $x\sim\tau(x)$. Then, for instance, arcs $\breve{l}$ in $\breve{X}$ correspond to ``doubled arcs'' $\breve{l}\cup\tau(\breve{l})$ in $X$ which are exchanged by $\tau$, and conversely. However, it is harder to ``properly divide'' a $\tau$-stable arc that encounters fixed points --- the latter are ``double counted'' in the na\"{i}ve quotient. It is then necessary to use an auxiliary construction, called the \emph{homotopy quotient} (or \emph{Borel construction}), in place of $\breve{X}$ --- this is defined to the the quotient E$\ZZ_2\times_{\ZZ_2} X$ where E$\ZZ_2$ is the \emph{universal $\ZZ_2$ bundle} over the \emph{classifying space} B$\ZZ_2$ and the quotient is by the \emph{free} diagonal $\ZZ_2$ action on E$\ZZ_2\times X$.

\begin{figure}[bth]
    \centering
\subfigure{
\begin{tikzpicture}[xscale=1, yscale=1,every node/.style={scale=0.7}]

\draw (0,0) -- (1,0) -- (1,2) -- (0,2) -- (0,0);
\draw (0,2) -- (0.8,2.6) -- (1.8,2.6) -- (1,2);
\draw (1,0) -- (1.8,0.6) -- (1.8,2.6);
\draw[dotted] (1,1) -- (1.8,1.6);
\draw[->] (0,2) -- (0,2.2);
\draw[->] (1,0) -- (1.2,0);
\draw[->] (0,0) -- (0.2,0.15);

\draw[dotted] (0,1) -- (1,1);
\draw[dotted] (0.8,1.6) -- (1.8,1.6);

\fill[gray!20,opacity=0.6] (1,1) -- (1.4,1.3) -- (1.4,2.3) -- (1,2) -- (1,1);
\fill[blue!20,opacity=0.6] (1.8,1.6) -- (1.4,1.3) -- (1.4,2.3) -- (1.8,2.6) -- (1.8,1.6);
\fill[blue!20,opacity=0.6] (1,1) -- (1.4,1.3) -- (1.4,0.3) -- (1,0) -- (1,1);
\fill[gray!20,opacity=0.6] (1.8,1.6) -- (1.4,1.3) -- (1.4,0.3) -- (1.8,0.6) -- (1.8,1.6);

\fill[purple!20,opacity=0.6] (0,1) -- (0.4,1.3) -- (0.4,2.3) -- (0,2) -- (0,1);
\fill[green!20,opacity=0.6] (0.8,1.6) -- (0.4,1.3) -- (0.4,2.3) -- (0.8,2.6) -- (0.8,1.6);
\fill[green!20,opacity=0.6] (0,1) -- (0.4,1.3) -- (0.4,0.3) -- (0,0) -- (0,1);
\fill[purple!20,opacity=0.6] (0.8,1.6) -- (0.4,1.3) -- (0.4,0.3) -- (0.8,0.6) -- (0.8,1.6);

\node[above] at (0,2.2) {$k_z$};
\node[right] at (1.2,0) {$k_x$};
\node[right] at (0.2,0.15) {$k_y$};

\node[below] at (0,0) {$-\pi$};
\node[left] at (0,1) {$0$};
\node[left] at (0,2) {$\pi$};
\node[below] at (1,0) {$0$};

\node at (1.4,1.3) [circle,fill=black]{} ;
\node at (1.8,1.6) [circle,fill=black]{} ;
\node at (1.4,2.3) [circle,fill=black]{} ;
\node at (1.8,2.6) [circle,fill=black]{} ;

\draw [red,thick,->-] (0.5,1) -- (1.3,1.6);

\end{tikzpicture}
}
\subfigure{
\begin{tikzpicture}[scale=1,every node/.style={scale=0.7}]

\draw (-0.7,0) -- (0,0);
\draw (-0.7,2) -- (0,2);
\draw (0.1,2.6) -- (0.8,2.6);
\draw (0,0) -- (0,2) -- (0.8,2.6) -- (0.8,0.6) -- (0,0);
\draw[dotted] (0,1) -- (0.8,1.6);
\draw[dotted] (-0.7,1) -- (0,1);
\draw[dotted] (0.1,1.6) -- (0.8,1.6);

\node at (0.4,1.3) [circle,fill=black]{} ;
\node at (0.8,1.6) [circle,fill=black]{};
\node at (0.4,2.3) [circle,fill=black]{} ;
\node at (0.8,2.6) [circle,fill=black]{};

\draw [red,thick,->-=0.7] (-0.5,1) .. controls (-0.1,1.4) .. (0.3,1.6);
\draw [red,thick,->-] (-0.5,1) -- (0.3,1.6);

\end{tikzpicture}
}
\hspace{-0.5em}
\subfigure{
\begin{tikzpicture}[scale=1,every node/.style={scale=0.7}]

\draw (-0.7,0) -- (0,0);
\draw (-0.7,2) -- (0,2);
\draw (0.1,2.6) -- (0.8,2.6);
\draw (0,0) -- (0,2) -- (0.8,2.6) -- (0.8,0.6) -- (0,0);
\draw[dotted] (0,1) -- (0.8,1.6);
\draw[dotted] (-0.7,1) -- (0,1);
\draw[dotted] (0.1,1.6) -- (0.8,1.6);

\fill[gray!20,opacity=0.3] (0,1) -- (0.4,1.3) -- (0.4,2.3) -- (0,2) -- (0,1);
\fill[blue!20,opacity=0.3] (0.8,1.6) -- (0.4,1.3) -- (0.4,2.3) -- (0.8,2.6) -- (0.8,1.6);
\fill[blue!20,opacity=0.3] (0,1) -- (0.4,1.3) -- (0.4,0.3) -- (0,0) -- (0,1);
\fill[gray!20,opacity=0.3] (0.8,1.6) -- (0.4,1.3) -- (0.4,0.3) -- (0.8,0.6) -- (0.8,1.6);

\node at (0.4,1.3) [circle,fill=black]{} ;
\node at (0.8,1.6) [circle,fill=black]{};
\node at (0.4,2.3) [circle,fill=black]{} ;
\node at (0.8,2.6) [circle,fill=black]{};

\draw [red,thick,->-=0.3] (-0.5,1) .. controls (-0.1,1.4) .. (0.3,1.6);
\draw [red,thick,->-=0.7] (0,1.5) -- (0.8,2.1);

\end{tikzpicture}
}
\hspace{-0.5em}
\subfigure{
\begin{tikzpicture}[scale=1,every node/.style={scale=0.7}]

\draw (-0.7,0) -- (0,0);
\draw (-0.7,2) -- (0,2);
\draw (0.1,2.6) -- (0.8,2.6);
\draw (0,0) -- (0,2) -- (0.8,2.6) -- (0.8,0.6) -- (0,0);
\draw[dotted] (0,1) -- (0.8,1.6);
\draw[dotted] (-0.7,1) -- (0,1);
\draw[dotted] (0.1,1.6) -- (0.8,1.6);

\fill[gray!20,opacity=0.3] (0,1) -- (0.4,1.3) -- (0.4,2.3) -- (0,2) -- (0,1);
\fill[blue!20,opacity=0.3] (0.8,1.6) -- (0.4,1.3) -- (0.4,2.3) -- (0.8,2.6) -- (0.8,1.6);
\fill[blue!20,opacity=0.3] (0,1) -- (0.4,1.3) -- (0.4,0.3) -- (0,0) -- (0,1);
\fill[gray!20,opacity=0.3] (0.8,1.6) -- (0.4,1.3) -- (0.4,0.3) -- (0.8,0.6) -- (0.8,1.6);

\node at (0.4,1.3) [circle,fill=black]{} ;
\node at (0.8,1.6) [circle,fill=black]{};
\node at (0.4,2.3) [circle,fill=black]{} ;
\node at (0.8,2.6) [circle,fill=black]{};

\draw [red,thick,->-=0.3] (-0.5,1) .. controls (-0.1,1.4) .. (0.3,1.6);
\draw [red,thick,->-] (0.8,1.1) -- (0,0.5);

\end{tikzpicture}
}
\hspace{-0.5em}
\subfigure{
\begin{tikzpicture}[scale=1,every node/.style={scale=0.7}]

\draw (-0.7,0) -- (0,0);
\draw (-0.7,2) -- (0,2);
\draw (0.1,2.6) -- (0.8,2.6);
\draw (0,0) -- (0,2) -- (0.8,2.6) -- (0.8,0.6) -- (0,0);
\draw[dotted] (0,1) -- (0.8,1.6);
\draw[dotted] (-0.7,1) -- (0,1);
\draw[dotted] (0.1,1.6) -- (0.8,1.6);

\node at (0.4,1.3) [circle,fill=black]{} ;
\node at (0.8,1.6) [circle,fill=black]{};
\node at (0.4,2.3) [circle,fill=black]{} ;
\node at (0.8,2.6) [circle,fill=black]{};

\draw [red,thick,->-=0.7] (-0.5,1) .. controls (-0.1,1.4) .. (0.3,1.6);
\draw [red,thick,->-=0.7] (0.3,1.6) .. controls (0,1.3) .. (-0.5,1);

\end{tikzpicture}
}
\hspace{-0.5em}
\subfigure{
\begin{tikzpicture}[scale=1,every node/.style={scale=0.7}]

\draw (-0.7,0) -- (0,0);
\draw (-0.7,2) -- (0,2);
\draw (0.1,2.6) -- (0.8,2.6);
\draw (0,0) -- (0,2) -- (0.8,2.6) -- (0.8,0.6) -- (0,0);
\draw[dotted] (0,1) -- (0.8,1.6);
\draw[dotted] (-0.7,1) -- (0,1);
\draw[dotted] (0.1,1.6) -- (0.8,1.6);

\node at (0.4,1.3) [circle,fill=black]{} ;
\node at (0.8,1.6) [circle,fill=black]{};
\node at (0.4,2.3) [circle,fill=black]{} ;
\node at (0.8,2.6) [circle,fill=black]{};

\draw [red,thick,->-=0.8] (-0.4,1.1) .. controls (-0.2,1.4) .. (0.2,1.5);
\draw [red,thick,->-=0.8] (0.2,1.5) .. controls (0.1,1.3) .. (-0.4,1.1);
\end{tikzpicture}
}

\caption{(Left) The red loop $\breve{l}_y$ in $\breve{\TT}^3$ is ``half'' of $l_y$ (the dual arcs to $\nu_y$). Starting from $2\breve{l}_y$, deform one $\breve{l}_y$ into a loop on the $k_x=0$ face, identify it with its $\tau$-image, then deform it back to $\breve{l}_y^{\rm op}$. Thus $2\breve{l}_y\sim\breve{l}_y+\breve{l}_y^{\rm op}$ which is contractible in $\breve{\TT}^3$.}\label{fig:torsionloops2}
\end{figure}
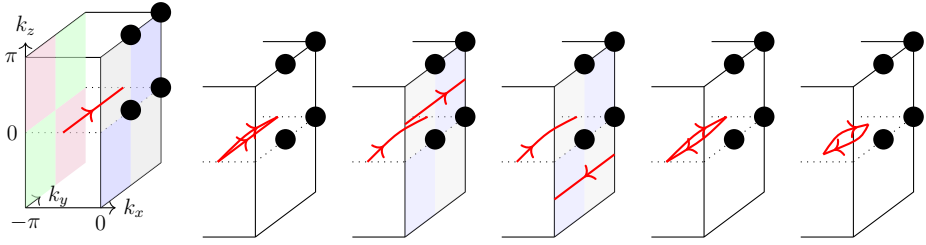

In Fig.\ \ref{fig:torsionloops2}, the quotient space $\breve{\TT}^3$ for the BZ is illustrated as an EBZ, where further $k\sim-k$ identifications are needed at the $k_x=-\pi,0$ faces. We also exhibit a loop in $\breve{\TT}^3$ with the property that looping twice is contractible.

\subsection{Topological bulk-boundary correspondence}\label{sec:mathbbc}
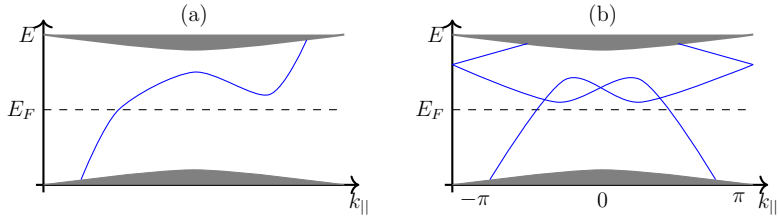
\begin{figure}
\centering
\subfigure{
\begin{tikzpicture}[xscale=1, every node/.style={scale=0.7}]
\draw[thick,->] (0,-0.1) -- (0,2.2);
\draw[thick,->] (-0.1,0) -- (4.2,0);
\node [below] at (4.2,0) {$k_{||}$};
\node [left] at (0,2) {$E$};
\node [left] at (0,1) {$E_F$};
\node [above] at (2,2) {(a)};

\draw[dashed] (0,1) -- (4,1);

\draw[gray, fill=gray] plot [smooth] coordinates {(0,0) (2,0.2) (4,0)};
\draw[gray, thick, fill=gray] plot [smooth] coordinates {(0,2) (2,1.8) (4,2)};

\draw[blue] plot [smooth] coordinates {(0.5,0.07) (1,1) (2,1.5) (3,1.2) (3.5,1.93)};

\end{tikzpicture}
}
\subfigure{
\begin{tikzpicture}[xscale=1, every node/.style={scale=0.7}]
\draw[thick,->] (0,-0.1) -- (0,2.2);
\draw[thick,->] (-0.1,0) -- (4.2,0);
\node [below] at (4.2,0) {$k_{||}$};
\node [left] at (0,2) {$E$};
\node [left] at (0,1) {$E_F$};
\node [below] at (2,0) {$0$};
\node [below right] at (0,0) {$-\pi$};
\node [below left] at (4,0) {$\pi$};
\node [above] at (2,2) {(b)};

\draw[dashed] (0,1) -- (4,1);

\draw[gray, fill=gray] plot [smooth] coordinates {(0,0) (2,0.2) (4,0)};
\draw[gray, thick, fill=gray] plot [smooth] coordinates {(0,2) (2,1.8) (4,2)};

\draw[blue] plot [smooth] coordinates {(0.5,0.07) (1.5,1.4) (2.5,1.1) (4,1.6)};
\draw[blue] plot [smooth] coordinates {(3.5,0.07) (2.5,1.4) (1.4,1.1) (0,1.6)};
\draw[blue] plot [smooth] coordinates {(0,1.6) (1,1.88)};
\draw[blue] plot [smooth] coordinates {(4,1.6) (3,1.88)};

\end{tikzpicture}
}
\caption{(a) Family of edge states (blue) connecting bulk bands (gray) in broken-$\mathsf{T}$ case. The number of intersections (chiral edge states) at the Fermi level $E_F$, counted with signs, stays invariant as $E_F$ is varied in the bulk gap. (b) Edge states in $\mathsf{T}$-symmetric case. Although the signed intersection number at $E_F$ always vanishes, the number of pairs of intersections (edge Kramers pairs) mod 2 is a $\ZZ_2$-invariant.}\label{fig:spectralflow}
\end{figure}

In the presence of a boundary, only the momenta $k_{||}$ parallel to the boundary remain well-defined. A bulk Hamiltonian $H$ with a spectral gap at the Fermi level $E_F$ has a spectral projection $p_F$ to the occupied states which defines the bulk invariant; partial Fourier-transform gives $H(k_{||})$ and a family of projections $p_F(k_{||})$. With boundary conditions imposed, $H(k_{||})$ becomes a half-space operator $\wt{H}(k_{||})$, whose corresponding $\wt{p}_F(k_{||})$ need not remain a projection. Intuitively, this means that new boundary-localised spectra with some $k_{||}$ is acquired at $E_F$. Since $E_F$ may be varied within the bulk gap, one sees that the boundary spectra actually fills up this gap and connects a bulk valence band to a conduction band. In 2D, $\wt{H}(k_{||})$ is just a periodic family over the edge momentum $k_{||}$, and we can ask how many times the edge spectra ``flows'' past the Fermi level from below as the $k_{||}$-circle is traversed around once (Fig.\ \ref{fig:spectralflow}), c.f.\ the Thouless charge pump. Such edge indices (Bott--Maslov/Toeplitz indices), spectral flow \cite{Hatsugai}, and the connection to the bulk topological invariants were studied for the 2D case in \cite{GP,ASV}, for the quantum Hall effect in \cite{KSB}, and more generally using a generalised cohomology theory called $K$-theory in \cite{PSB,MT2}. 

The (Toeplitz) bulk-boundary map of \cite{KSB} is formulated as a certain homomorphism $\delta$ between bulk and boundary topological invariants. Abstractly, $\delta$ is a kind of \emph{push-forward/Gysin map} associated to the projection $k\mapsto k_{||}$, which in the broken-$\mathsf{T}$ case is just integration along the transverse momenta $k_\perp$ \cite{BottTu}. Concretely, $\delta(p_F)$ measures topologically the failure of $\wt{p}_F$ to remain a projection, and can be interpreted as counting \emph{spectral flow}. In the $\mathsf{T}$-symmetric case, the bulk-boundary map can be analogously defined to be ``$\mathsf{T}$-invariant integration'' $\mathcal{P}$, as in the main text. The precise interpretation of $\mathcal{P}$ as spectral flow in this case, especially in higher dimensions for semimetals, is not so well-understood and is the subject of ongoing investigation \cite{GT}. Heuristically, the bulk band topology determines topological features of the boundary spectra. The latter is probed by the boundary Fermi surface which stays in the same topological class as the Fermi energy is varied within the bulk gap. Since the bulk band topology is classified cohomologically (e.g.\ through $\mathcal{H}^2$), it is reasonable to expect the same for the boundary Fermi surface, at least in a dual sense (homologically, e.g.\ by $\mathcal{H}_1$).

\subsection{Poincar\'{e} duality, integration, and intersection numbers}\label{sec:mathpoincare}
If $X$ is a closed oriented $d$-dimensional manifold, there is a canonical isomorphism $H^{n}(X)\cong H_{d-n}(X)$ for each $n$, known as Poincar\'{e} duality (PD) \cite{BottTu,Hatcher}. We are mainly concerned with $d=3$, so $H^2(X)$ is, under PD, the same thing as homology classes of arcs, $H_1(X)$. If $M\subset X$ is a closed oriented $1$-dimensional submanifold representing a class in $H_1(X)$, its PD in $H^2(X)$ is (the cohomology class of) a 2-form $\omega$ satisfying $\int_X\omega\wedge\eta=\int_M\eta$ for \emph{all closed} 1-forms $\eta$. Intuitively, such an $\omega$ ``sees'' all the directions normal to $M$. Another geometrical interpretation is that the PD of a $(d-n)$-cycle in a $d$-manifold $X$ is the $n$-cocycle that assigns to each $n$-cycle the number of (signed) intersections it has with the given $(d-n)$-cycle \cite{Hatcher}.

To illustrate the basic idea of PD in the context of topological insulators, consider a weak 3D Chern insulator built from layers of unit-Chern-number 2D insulators parallel to the $y$-$z$ directions. Its Berry curvature $\mathcal{F}$ is such that the integral over $k_y$-$k_z$ is the non-zero weak Chern number $c_1^{yz}=1$, while the integrals over $k_x$-$k_y$ and $k_x$-$k_z$ vanish, thus $\mathcal{F}$ is $dk_y\wedge dk_z$ up to some exact form and $2\pi$ normalization factors. Then $\mathcal{F}$ is PD to (any) circle $S^1_x$ wrapping $k_x$ once, since it is easy to verify that $\int_{\TT^3} \mathcal{F}\wedge\eta=\int_{S^1_x}\eta$ for all closed 1-forms $\eta$: we only need to check $\eta=f(k)dk_x$, and $d\eta=0$ ensures that $\int_{S^1_x}\eta$ does not depend on the $k_y,k_z$ coordinates of $S^1_x$. Thus this weak Chern insulator may be represented by the circle $S^1_x$ (at any $k_y,k_z$).

If we integrate $\mathcal{F}$ over $k_y$, we are left with a 1-form $dk_z$ on the surface BZ $\TT^2$, whose PD (in $\TT^2$) is now $S^1_x\subset \TT^2$ and turns out to be the surface Fermi circle. Thus ${\rm PD}(\int_{S^1_y} \mathcal{F})=p({\rm PD}(\mathcal{F}))$ where $p$ is the map projecting out $k_y$, i.e.\ $p$ is Poincar\'{e} dual to integrating over $k_y$. Furthermore, $S^1_x$ generically intersects (transversally in the BZ) the $k_y$-$k_z$ subtorus once but not the other subtori, and these intersection numbers recover the weak Chern numbers $c_1^{yz}=1, c_1^{xy}=0=c_1^{zx}$.

When there is a set of (Weyl) points $W$ to be excluded from $X$, PD generalises (also called \emph{Poincar\'{e}--Lefschetz duality} \cite{Hatcher}) to $H^n(X\setminus W)\cong H_{d-n}(X,W)$, utilised to analyse WSM in \cite{MT,MT1}. If $X$ has a smooth $\ZZ_2$-action $\tau$, with some minor assumptions on $(X,\tau)$ satisfied by all examples in this paper, an equivariant version of PD holds for $\mathcal{H}^\bullet$ and $\mathcal{H}_\bullet$ \cite{GT}. For example, there are isomorphisms $\mathcal{H}^n(\TT^3)\cong \mathcal{H}_{3-n}(\TT^3)$ and $\mathcal{H}^n(\TT^3\setminus W)\cong \mathcal{H}_{3-n}(\TT^3,W)$ for $n\geq 2$ and $W$ a finite set of points avoiding the TRIM set $F$.

\subsection{Global invariants for $\Twsm$ and Weyl charge cancellation}\label{sec:mathMV}
For 2D/3D TI, the $\Tinv$ valence bands can always be trivialised (global choice of Bloch eigenstates spanning the valence states exists), and one can prove that $\mathcal{H}^2(\TT^d), d=2,3$ ``localises'' completely to data at the TRIM. Explicitly, this means that $\mathcal{H}^2(\TT^d)$ is given by sign maps $F\rightarrow\{\pm\}$ modulo global $\mathsf{T}$-gauge transformations over $\TT^d$, recovering the Fu--Kane--Mele POS formulae for the FKMI \cite{dNG}. However, for 3D $\Twsm$, it is not possible to make a global choice of valence Bloch eigenstates everywhere on $\TT^3\setminus W$, so signs cannot be assigned to all the TRIM at once. Nevertheless, we can still compute the group $\mathcal{H}^2(\TT^3\setminus W)\cong\mathcal{H}_1(\TT^3,W)$ of $\Twsm$ invariants using a general ``locality principle''. This is called the \emph{Mayer--Vietoris} property of (co)homology \cite{BottTu}, which is a kind of inclusion-exclusion principle that allows topological invariants of $X$ to be computed from (i) those of $\Tinv$-subspaces $\{X_i\}_{i\in I}$ that cover $X$, and (ii) those of their intersections.

Consider the simplest $\Twsm$ with two pairs of Weyl points $W=\{\pm w,\pm w'\}$ disjoint from the TRIM, as in the main text. Note that $w$ and $-w$ have the same Weyl charge by $\mathsf{T}$-symmetry, as do $w'$ and $-w'$. Take $X_1=\TT^3\setminus W$ and $X_2=D_W$ where $D_W$ is the ($\Tinv$) disjoint union of small open balls surrounding the Weyl points, then $X_1\cap X_2$ is (retractable to) $S^2_W\equiv S^2_w\coprod S^2_{-w}\coprod S^2_{w'}\coprod S^2_{-w'}$. The Mayer--Vietoris principle organises the $\mathcal{H}^\bullet,\mathcal{H}_\bullet$ of all these spaces, by producing the sequence (c.f.~\cite{MT})

\begin{equation*}
\footnotesize
    \xymatrix{ \cdots\rightarrow 0\ar[r]&{\overbrace{\mathcal{H}^2(\TT^3)}^{\substack{{\rm TI} \\ {\rm invariants}}}}\ar[d]_{\text{PD}}\ar[r]^{\rm restrict\;\;\;\;}_{\rm bands\;\;\;\;} & {\overbrace{\mathcal{H}^{2}(\TT^3\setminus W)}^{\substack{\Twsm \\ {\rm invariants}}}}\ar[d]_{\text{PD}}\ar[r] ^{\rm \;\; restrict}_{\rm \;\; bands}& {\overbrace{\mathcal{H}^{2}(S^2_W)}^{\substack{\Twsm\,{\rm local} \\ {\rm Weyl\,charges}}}}\ar[d]_{\text{PD}}\ar[r]^{\rm sum} & {\overbrace{\mathcal{H}^3(\TT^3)}^{\substack{{\rm total} \\ {\rm charge}}}}\ar[d]_{\text{PD}}\ar[r] & 0\rightarrow\cdots\\
   \cdots\rightarrow 0\ar[r] & {\underbrace{\mathcal{H}_1(\TT^3)}_{\substack{{\rm Dual\,TI} \\ {\rm arcs}}}}\ar[r] & {\underbrace{\mathcal{H}_1(\TT^3,W)}_{\substack{{\rm Dual\, \Twsm} \\ {\rm arcs}}}} \ar[r]^{\partial={\rm local}}_{\rm charges} &  \mathcal{H}_0(W) \ar[r] & \mathcal{H}_0(\TT^3)\ar[r] & 0\rightarrow\cdots
   \normalsize
    }
\end{equation*}
in which both the horizontal sequences are exact (i.e.\ the kernel of one map \emph{equals} the image of the previous map) and the vertical maps are Poincar\'{e} duality isomorphisms. Note that $D_W$ does not contribute because it is contractible to (the Weyl) points. The local Weyl charge group for $\Twsm$ is $\ZZ^2$, given by the Weyl charges of $w$ and $w'$.

Exactness is precisely the useful property for computation and physical interpretation. For example, exactness at $\mathcal{H}^2(S^2_W)$ means that when a $\Twsm$ is restricted to the spheres $S^2_W$ surrounding the Weyl points (image of the second restriction map), the local charges must cancel (kernel of the sum map), and furthermore, every charge-cancelling configuration can arise in some $\Twsm$. This is the precise charge-cancellation condition for $\Twsm$, included here because the oft-cited Nielsen--Ninomiya argument (used for ordinary WSM) does not work verbatim. 

The top sequence is, explicitly,
\begin{equation}
0\rightarrow\ZZ_2^4\rightarrow \mathcal{H}^{2}(\TT^3\setminus W)\cong\mathcal{H}_1(\TT^3,W)\overset{\rm local\, charges}{\longrightarrow} \ZZ^2\overset{\text{sum}}{\rightarrow} \ZZ\rightarrow 0,\label{homologysequence}
\end{equation}
from which we deduce that $\mathcal{H}^{2}(\TT^3\setminus W)\cong\ZZ_2^4\oplus\ZZ\cong\mathcal{H}_1(\TT^3,W)$. The $\ZZ_2^4$ factors are the usual TI invariants $\nu_0,\nu_i$, while the new integer invariant $q\in\ZZ$ counts the number of (reference) ``open arcs'' connecting $\pm w'$ to $\pm w$. As in the main text, a pair of open arcs connecting $\pm w'$ to $\mp w$ represents $(\nu_0;\nu_i;q)=(-;+,+,+;1)$, and generates a \emph{different} copy $\ZZ'$ of $\ZZ$ in $\mathcal{H}_1(\TT^3,W)$. Thus a particular choice of reference Weyl point connection just selects a $\ZZ$ basis element for the $\Twsm$ invariants (i.e.\ it specifies one of many possible factorisations of $\mathcal{H}_1(\TT^3,W)$ into $\ZZ_2^4\oplus\ZZ$). In group-theoretic terms, the abstract abelian group $\mathcal{H}_1(\TT^3,W)$ has a \emph{torsion subgroup} $T=\ZZ_2^4$ (the subset of elements of finite order) while the \emph{quotient group} $F\cong\ZZ$ is torsion-free. The splitting $\mathcal{H}_1(\TT^3,W)\cong T\oplus F=\ZZ_2^4\oplus\ZZ$ into a direct sum is not canonical, but depends on a choice of the lift of $F$ into $\mathcal{H}_1(\TT^3,W)$. Two possible lifts of the generator of $F$ were specified above, namely $(+;+,+,+;1)$ and $(-;+,+,+;1)$, and they lead to two different ways of writing $\mathcal{H}_1(\TT^3,W)\cong\ZZ_2^4\oplus\ZZ\cong \ZZ_2^4\oplus'\ZZ'$, related by a non-trivial automorphism mapping $(+;+,+,+;1)$ to $(-;+,+,+;1)$.

\subsection{Generalisation to more pairs of Weyl points}\label{section:morepairs}
If there are $r$ pairs of Weyl points, $W=\{\pm w_i\}_{i=1,\ldots,r}$, we will have $\mathcal{H}_1(\TT^3,W)\cong\ZZ_2^4\oplus\ZZ^{r-1}$, where the factorisation again depends on a choice of reference basis connections. A natural choice is to take, for each $i=1,\ldots,r-1$, a $\Tinv$ pair of connections, denoted $l_{i+1,i}$, from $\pm w_{i+1}\rightarrow\pm w_i$ as the generator of a $\ZZ$. Let $q_i\in\ZZ$ denote the common local Weyl charge of $\pm w_i$, necessarily satisfying $q_r=-\sum_{1\leq j\leq r-1}q_j$. Without loss of generality\footnote{After possibly adding some $\Tinv$ loops such as $\pm w_i\rightarrow\pm w_j\rightarrow \pm w_k\rightarrow \pm w_{i}$ (trivial loops), or $w_i\rightarrow-w_j\rightarrow-w_i\rightarrow w_j\rightarrow w_i$ (representing $\nu_0$), or $l_x,l_y,l_z$ (representing $\nu_x,\nu_y,\nu_z$.)}, we can assume that a Weyl point history is given by a collection of $\Tinv$ loops/arcs in which the only connections are $l_{i+1,i}$ (in particular, there are no connections between $\pm w_i$ and $\mp w_j$, or between $\pm w_1$ and $\pm w_r$). The number of reference arcs $l_{i+1,i}$ entering $w_i$ (or $-w_i$) minus the number of reference arcs $l_{i,i-1}$ leaving $w_i$ (or $-w_i$) must equal $q_i$ (for $i=1$ there are no reference arcs leaving $w_1$). A telescoping sum shows that the number of reference $l_{i+1,i}$ is given by the integer $q_1+\ldots+q_i$, so the topological numbers for the $\Twsm$ can be taken to be $$\boldsymbol{\nu}\equiv(\nu_0;\nu_x,\nu_y,\nu_z;q_1, q_1+q_2, \ldots,\sum_{1\leq j\leq r-1}q_j)\in\ZZ_2^4\oplus\ZZ^{r-1}.$$

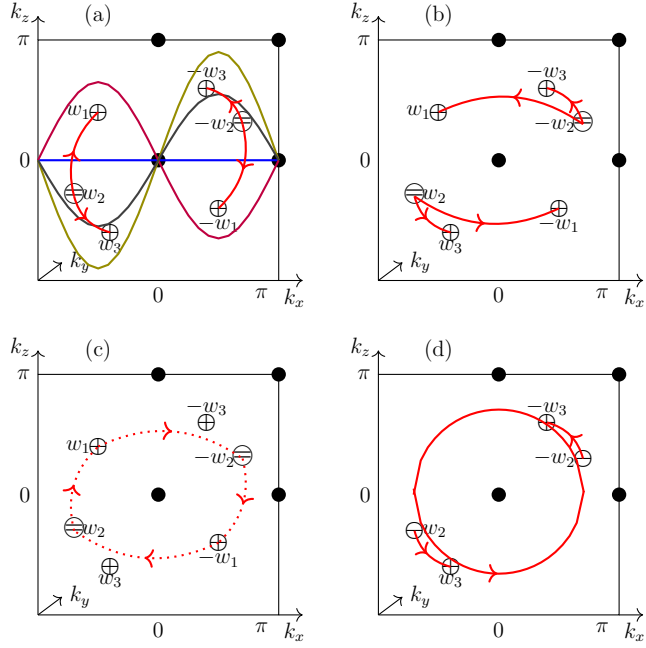
\begin{figure}
\centering
\subfigure{
\begin{tikzpicture}[scale=1.6, every node/.style={scale=0.7}]

\node at (0.5,2.2) {(a)};

\draw (0,0) -- (0,2) -- (2,2) -- (2,0) -- (0,0);
\draw[->] (0,2) -- (0,2.2);
\draw[->] (2,0) -- (2.2,0);
\draw[->] (0,0) -- (0.2,0.15);

\node [right] at (0.2,0.15) {$k_y$};
\node [left] at (0,2.23) {$k_z$} ;
\node [below] at (2.13,0) {$k_x$} ;
\node [below] at (1.86,0) {$\pi$} ;
\node [left] at (0,2) {$\pi$} ;
\node [left] at (0,1) {$0$};
\node [below] at (1,0) {$0$};

\node at (1,1) [circle,fill=black,scale=0.7]{} ;
\node at (2,1) [circle,fill=black,scale=0.7]{} ;
\node at (2,2) [circle,fill=black,scale=0.7]{} ;
\node at (1,2) [circle,fill=black,scale=0.7]{} ;

\node at (0.3,0.7) {$\boldsymbol{\oeq}$} ;
\node at (0.5,1.4) {$\boldsymbol{\oplus}$} ;
\node at (1.5,0.6) {$\boldsymbol{\oplus}$} ;
\node at (1.7,1.3) {$\boldsymbol{\oeq}$} ;
\node at (1.4,1.6) {$\boldsymbol{\oplus}$} ;
\node at (0.6,0.4) {$\boldsymbol{\oplus}$} ;

\node[right] at (0.3,0.7) {$w_2$} ;
\node[left] at (0.5,1.4) {$w_1$} ;
\node[below] at (1.5,0.6) {$-w_1$} ;
\node[left] at (1.7,1.3) {$-w_2$} ;
\node[above] at (1.4,1.6) {$-w_3$} ;
\node[below] at (0.6,0.4) {$w_3$} ;

\draw[thick,blue,domain=0:2] plot (\x, 1);
\draw[thick,purple,domain=0:2] plot (\x, {1+0.65*sin(\x*pi r) });
\draw[thick,olive,domain=0:2] plot (\x, {1-0.9*sin(\x*pi r) });
\draw[thick,darkgray,domain=0:2] plot (\x, {1-0.55*sin(\x*pi r) });

\draw[->-] [red,thick] (0.3,0.7) to [bend left] (0.5,1.4);
\draw[->-] [red,thick] (1.7,1.3) to [bend left] (1.5,0.6);
\draw[->-] [red,thick] (1.7,1.3) to [bend right] (1.4,1.6);
\draw[->-] [red,thick] (0.3,0.7) to [bend right] (0.6,0.4);

\end{tikzpicture}
}
\subfigure{
\begin{tikzpicture}[scale=1.6, every node/.style={scale=0.7}]

\node at (0.5,2.2) {(b)};

\draw (0,0) -- (0,2) -- (2,2) -- (2,0) -- (0,0);
\draw[->] (0,2) -- (0,2.2);
\draw[->] (2,0) -- (2.2,0);
\draw[->] (0,0) -- (0.2,0.15);

\node [right] at (0.2,0.15) {$k_y$};
\node [left] at (0,2.23) {$k_z$} ;
\node [below] at (2.13,0) {$k_x$} ;
\node [below] at (1.86,0) {$\pi$} ;
\node [left] at (0,2) {$\pi$} ;
\node [left] at (0,1) {$0$};
\node [below] at (1,0) {$0$};

\node at (1,1) [circle,fill=black,scale=0.7]{} ;
\node at (2,1) [circle,fill=black,scale=0.7]{} ;
\node at (2,2) [circle,fill=black,scale=0.7]{} ;
\node at (1,2) [circle,fill=black,scale=0.7]{} ;

\node at (0.3,0.7) {$\boldsymbol{\oeq}$} ;
\node at (0.5,1.4) {$\boldsymbol{\oplus}$} ;
\node at (1.5,0.6) {$\boldsymbol{\oplus}$} ;
\node at (1.7,1.3) {$\boldsymbol{\oeq}$} ;
\node at (1.4,1.6) {$\boldsymbol{\oplus}$} ;
\node at (0.6,0.4) {$\boldsymbol{\oplus}$} ;

\node[right] at (0.3,0.7) {$w_2$} ;
\node[left] at (0.5,1.4) {$w_1$} ;
\node[below] at (1.5,0.6) {$-w_1$} ;
\node[left] at (1.7,1.3) {$-w_2$} ;
\node[above] at (1.4,1.6) {$-w_3$} ;
\node[below] at (0.6,0.4) {$w_3$} ;

\draw[->-] [red,thick] (1.7,1.3) to [bend right] (0.5,1.4);
\draw[->-] [red,thick] (0.3,0.7) to [bend right] (1.5,0.6);
\draw[->-] [red,thick] (1.7,1.3) to [bend right] (1.4,1.6);
\draw[->-] [red,thick] (0.3,0.7) to [bend right] (0.6,0.4);

\end{tikzpicture}
}
\hspace{2.5em}
\subfigure{
\begin{tikzpicture}[scale=1.6, every node/.style={scale=0.7}]

\node at (0.5,2.2) {(c)};

\draw (0,0) -- (0,2) -- (2,2) -- (2,0) -- (0,0);
\draw[->] (0,2) -- (0,2.2);
\draw[->] (2,0) -- (2.2,0);
\draw[->] (0,0) -- (0.2,0.15);

\node [right] at (0.2,0.15) {$k_y$};
\node [left] at (0,2.23) {$k_z$} ;
\node [below] at (2.13,0) {$k_x$} ;
\node [below] at (1.86,0) {$\pi$} ;
\node [left] at (0,2) {$\pi$} ;
\node [left] at (0,1) {$0$};
\node [below] at (1,0) {$0$};

\node at (1,1) [circle,fill=black,scale=0.7]{} ;
\node at (2,1) [circle,fill=black,scale=0.7]{} ;
\node at (2,2) [circle,fill=black,scale=0.7]{} ;
\node at (1,2) [circle,fill=black,scale=0.7]{} ;

\node at (0.3,0.7) {$\boldsymbol{\oeq}$} ;
\node at (0.5,1.4) {$\boldsymbol{\oplus}$} ;
\node at (1.5,0.6) {$\boldsymbol{\oplus}$} ;
\node at (1.7,1.3) {$\boldsymbol{\oeq}$} ;
\node at (1.4,1.6) {$\boldsymbol{\oplus}$} ;
\node at (0.6,0.4) {$\boldsymbol{\oplus}$} ;

\node[right] at (0.3,0.7) {$w_2$} ;
\node[left] at (0.5,1.4) {$w_1$} ;
\node[below] at (1.5,0.6) {$-w_1$} ;
\node[left] at (1.7,1.3) {$-w_2$} ;
\node[above] at (1.4,1.6) {$-w_3$} ;
\node[below] at (0.6,0.4) {$w_3$} ;

\draw[->-] [red,dotted,thick] (0.5,1.4) to [bend left] (1.7,1.3);
\draw[->-] [red,dotted,thick] (1.5,0.6) to [bend left] (0.3,0.7);
\draw[->-] [red,dotted,thick] (0.3,0.7) to [bend left] (0.5,1.4);
\draw[->-] [red,dotted,thick] (1.7,1.3) to [bend left] (1.5,0.6);

\end{tikzpicture}
}
\subfigure{
\begin{tikzpicture}[scale=1.6, every node/.style={scale=0.7}]

\node at (0.5,2.2) {(d)};

\draw (0,0) -- (0,2) -- (2,2) -- (2,0) -- (0,0);
\draw[->] (0,2) -- (0,2.2);
\draw[->] (2,0) -- (2.2,0);
\draw[->] (0,0) -- (0.2,0.15);

\node [right] at (0.2,0.15) {$k_y$};
\node [left] at (0,2.23) {$k_z$} ;
\node [below] at (2.13,0) {$k_x$} ;
\node [below] at (1.86,0) {$\pi$} ;
\node [left] at (0,2) {$\pi$} ;
\node [left] at (0,1) {$0$};
\node [below] at (1,0) {$0$};

\node at (1,1) [circle,fill=black,scale=0.7]{} ;
\node at (2,1) [circle,fill=black,scale=0.7]{} ;
\node at (2,2) [circle,fill=black,scale=0.7]{} ;
\node at (1,2) [circle,fill=black,scale=0.7]{} ;

\node at (0.3,0.7) {$\boldsymbol{\ominus}$} ;
\node at (1.7,1.3) {$\boldsymbol{\ominus}$} ;
\node at (1.4,1.6) {$\boldsymbol{\oplus}$} ;
\node at (0.6,0.4) {$\boldsymbol{\oplus}$} ;

\node[right] at (0.3,0.7) {$w_2$} ;
\node[left] at (1.7,1.3) {$-w_2$} ;
\node[above] at (1.4,1.6) {$-w_3$} ;
\node[below] at (0.6,0.4) {$w_3$} ;

\draw [red,thick,domain=-0.7071:0.70711] plot (1+\x, {1+(0.5-\x*\x)^(0.5) });
\draw [red,thick,->-,domain=-0.7071:0.70711] plot (1+\x, {1.05-(0.5-\x*\x)^(0.5) });

\draw[->-] [red,thick] (1.7,1.3) to [bend right] (1.4,1.6);
\draw[->-] [red,thick] (0.3,0.7) to [bend right] (0.6,0.4);

\end{tikzpicture}
}
\caption{The $\oeq$ symbol indicates that the Weyl points $\pm w_2$ each has charge $q_2=-2$. Weyl point histories are represented by red directed arcs. For Fig.\ \ref{fig:multiweyl}(a), we deduce in the same way as Eq.\ \eqref{blue-purple} that $+=\nu_z\equiv\nu_{z,\pi}=-\nu_{z,0}=-\nu_{z,0}^{\rm \small blue}=\nu_{z,0}^{\rm purple}=-\nu_{z,0}^{\rm gray}=\nu_{z,0}^{\rm olive}$, which can also be deduced by counting mod 2 pairs of intersections of the respective curved $\Ttorus$ with the Weyl history. (b) A different set of connections between the same Weyl points. (c) Dotted red line represents $\nu_0=-$, which when added to (b) gives (a). The resultant history starting from (b) and partially annihilating Weyl points vertically is depicted in (d).}\label{fig:multiweyl}
\end{figure}

Let us illustrate the $r=3$ case in more detail. Equation \eqref{homologysequence} for $W=\{\pm w_1,\pm w_2, \pm w_3\}$ is now
\begin{equation*}
0\rightarrow\ZZ_2^4\rightarrow\mathcal{H}_1(\TT^3,W)\overset{\rm local\, charges}{\longrightarrow} \ZZ^3\overset{\text{sum}}{\rightarrow} \ZZ\rightarrow 0.
\end{equation*}
The local charge map takes a $\Twsm$ (arcs representing the Weyl point history) into $(q_1,q_2,q_3)\in\ZZ^3$, where $q_i$ is the Weyl charge of $\pm w_i$ and $q_3=-q_1-q_2$ necessarily. Then $\mathcal{H}_1(\TT^3,W)\cong\ZZ_2^4\oplus\ZZ^2$ where the $\ZZ^2$ factor is taken with respect to the reference connections $l_{2,1}$ and $l_{3,2}$. An example with $q_1=+1, q_2=-2$ (and $q_3=+1$), and with all three pairs of Weyl points in the $k_y=0$ $\Ttorus$ for simplicity, is shown in Fig.\ \ref{fig:multiweyl}(a). It has topological numbers $\boldsymbol{\nu}\equiv(\nu_0;\nu_i;q_1,q_1+q_2)=(+;+,+,+;1,-1)$ where the last two integers count the number of reference connections $l_{2,1}, l_{3,2}$ respectively. Consider another $\Twsm$ with the same Weyl points and local charges, but different history, as in Fig.\ \ref{fig:multiweyl}(b). By adding the $\mathsf{T}$-loop of Fig.\ \ref{fig:multiweyl}(c) (which represents a strong invariant), we recover the $\Twsm$ history of Fig.\ \ref{fig:multiweyl}(a). Thus the topological numbers for Fig.\ \ref{fig:multiweyl}(b) are $\boldsymbol{\nu}=(-;+,+,+;1,-1)$. An alternative way to see that $\nu_0=-$ for Fig.\ \ref{fig:multiweyl}(b) is to annihilate $w_1$ (resp.\ $-w_1$) with one of the $-1$ charges in $w_2$ (resp.\ $-w_2$) ``vertically'' along the arc $l_{2,1}$, to produce the history in Fig.\ \ref{fig:multiweyl}(d); upon further annihilation of the remaining $-1$ charge of $\pm w_2$ with $\pm w_3$ along the arc $l_{3,2}$, this history becomes a single loop representing the TI with $\nu_0=-$.

The difference in $\nu_0$ for Fig.\ \ref{fig:multiweyl}(a)-(b) also illustrates the dependence of the topological numbers on the choice of reference connections. Suppose we had taken $l'_{2,1}=\pm w_2\rightarrow\mp w_1$ and $l_{3,2}$ as references instead. There must still be $q_1$ and $q_1+q_2$ of these reference connections, respectively. The topological numbers under this alternative convention are $\boldsymbol{\nu}'\equiv(\nu'_0;\nu_i;q_1,q_1+q_2)$ where, as before, $\nu'_0$ is the strong invariant of the TI that results from annihilating Weyl points along the new reference connections. Then Fig.\ \ref{fig:multiweyl}(b) has $\boldsymbol{\nu}'=(+;+,+,+;1,-1)$, while Fig.\ \ref{fig:multiweyl}(a) has $\boldsymbol{\nu}'=(-;+,+,+;1,-1)$.

\section{Numerical methods: Spectral density for the interface layer}\label{sec:numerics}
In order to find localized modes between two topologically distinct systems, we used the method based on Green's function provided in \cite{Sancho, Lau}. Suppose that the Hamiltonian is given by
\begin{eqnarray*}
H&=&\sum_{n,m}\left(H_n\delta_{n,m}+V_n\delta_{n,m-1}+V_{n-1}^\dagger\delta_{n,m+1}\right)e_{n,m}\,,\\
&=&\sum_n\left( H_ne_{n,n}+V_ne_{n,n+1}+V_n^\dagger e_{n+1,n}\right)\,,
\end{eqnarray*}
where $n,m$ label layers, $H_n$ and $V_n$ are $M\times M$ matrices with $M$ being the number of internal degrees of freedom, and $e_{n,m}$ are shift operators from position $m$ to $n$. Let us consider the case in which system A is stacked on top of B. The corresponding Hamiltonian is described by $H_n=h_A(h_B)$ for $n<0$ ($n>0$), $H_0=\tilde h$, and $V_n=v_A(v_B)$ for $n\leq -1$ ($n\geq 0$). The Green's function is then defined by $\left(\omega I-H\right)G(\omega)=I$, where $I$ is the identity matrix corresponding to the dimension of $H$. Letting $G_{n,m}$ be $M\times M$ block part of the Green's function, which pertains to $n$-th and $m$-th layers, we find the following set of equations:
\begin{eqnarray*}
\left(\omega I_M-\tilde h\right)G_{0,0}&=&I_M+v_A^\dagger G_{-1,0}+v_BG_{1,0}\,,\nonumber\\
\left(\omega I_M-h_A\right)G_{n,0}&=&v_A^\dagger G_{n-1,0}+v_AG_{n+1,0}\,,\quad \text{for $n\leq -1$}\,,\\
\left(\omega I_M-h_B\right)G_{n,0}&=&v_B^\dagger G_{n-1,0}+v_BG_{n+1,0}\,,\quad \text{for $n\geq 1$}\,,\nonumber
\end{eqnarray*}
with $I_M$ being $M\times M$ identity matrix. The spectral density of the localized mode at the interface is related to $G_{0,0}$, which can be found by recursively solving the above set of equations. Defining the following relations for the first step in the recursion
\begin{eqnarray*}
\alpha_1^{A,B}&=&v_{A,B}\left(\omega I_M-h_{A,B}\right)^{-1}v_{A,B}\,,\\
\beta_1^{A,B}&=&v_{A,B}^\dagger\left(\omega I_M-h_{A,B}\right)^{-1}v_{A,B}^\dagger\,,\\
\kappa_1^{A,B}&=&v_{A,B}\left(\omega I_M-h_{A,B}\right)^{-1}v_{A,B}^\dagger\,,\\
\lambda_1^{A,B}&=&v_{A,B}^\dagger\left(\omega I_M-h_{A,B}\right)^{-1}v_{A,B}\,,\\
h_{0}^{A,B}&=&h_{A,B}\,,\qquad
h_{0}=\tilde h\,,
\end{eqnarray*}
the set of relations in $l$-th step are given by
\begin{eqnarray*}
h_{l}^{A,B}&=&h_{l-1}^{A,B}+\kappa_l^{A,B}+\lambda_l^{A,B}\,,\qquad
h_{l}=h_{l-1}+\kappa_l^{B}+\lambda_l^{A}\,,\\
\alpha_l^{A,B}&=&\alpha_{l-1}^{A,B}\left(\omega I_M-h_{l-1}^{A,B}\right)^{-1}\alpha_{l-1}^{A,B}\,,\\
\beta_l^{A,B}&=&\beta_{l-1}^{A,B}\left(\omega I_M-h_{l-1}^{A,B}\right)^{-1}\beta_{l-1}^{A,B}\,,\\
\kappa_l^{A,B}&=&\alpha_{l-1}^{A,B}\left(\omega I_M-h_{l-1}^{A,B}\right)^{-1}\beta_{l-1}^{A,B}\,,\\
\lambda_l^{A,B}&=&\beta_{l-1}^{A,B}\left(\omega I_M-h_{l-1}^{A,B}\right)^{-1}\alpha_{l-1}^{A,B}\,,
\end{eqnarray*}
and the set of the equations for the Green's function in $l$-th step are obtained as follows: 
\begin{eqnarray*}
\left(\omega I_M-h_l\right)G_{0,0}&=&I_M+\beta_l^AG_{-2^l,0}+\alpha_l^BG_{2^l,0}\,,\\
\left(\omega I_M-h^A_l\right)G_{n,0}
&=&\beta_l^AG_{n-2^l,0}
+\alpha_l^AG_{n+2^l,0}\,,\quad \text{for $n\leq -2^l$}\,,\\
\left(\omega I_M-h^B_l\right)G_{n,0}
&=&\beta_l^BG_{n-2^l,0}
+\alpha_l^BG_{n+2^l,0}\,,\quad \text{for $n\geq 2^l$}\,.
\end{eqnarray*}
Typically $\alpha_l^{A,B}$ and $\beta_l^{A,B}$ reduce in magnitude as $l$ increases, so we can approximate the Green's function at zeroth layer by $G_{0,0}(\omega)\approx\left(\omega I_M-h_l\right)^{-1}$. 
The spectral function at the zeroth layer is then given by $A_{0,0}(\omega)=-(2\pi)^{-1}{\rm Im}\left\{{\rm Tr}\left[G_{0,0}(\omega+i\delta)\right]\right\}$, where we let $\omega\rightarrow\omega+i\delta$ with $\delta$ being a sufficiently small real number.

For a concrete example giving Fig.\ \ref{fig:ABstack}(a)-(d) in the main text, we can choose the Hamiltonian for the system A to be
\begin{eqnarray*}
h_A(k_x,k_z)&=&a\sin k_x\tau_1\otimes\sigma_3+\beta\tau_2\otimes\sigma_2+d\tau_2\otimes\sigma_3+
\\
&{}&\,\,\left[t\cos k_z+2b\left(2-\cos k_x\right)\right]\tau_3\otimes\sigma_0+\lambda\sin k_z\tau_0\otimes\sigma_1,\\
v_A&=&b\tau_3\otimes\sigma_0+\frac{i}{2}\left(a\tau_2\otimes\sigma_0+\alpha\tau_1\otimes\sigma_2\right)\,,
\end{eqnarray*}
which are obtained from the tight-binding model for a time-reversal invariant Weyl semimetal given in \cite{Lau} by only Fourier transforming $x$- and $z$-directions. Here, $\tau_0$ and $\sigma_0$ are $2\times 2$ identity matrices, and $\tau_i$ and $\sigma_i$ are Pauli matrices. Furthermore, we set the system B by exchanging $k_x$ and $k_z$ and inverting the coordinate in $y$-direction from the system A, which translates to
\begin{eqnarray*}
h_B(k_x,k_z)&=&h_A(k_z,k_x)\,,\\
v_B&=&b\tau_3\otimes\sigma_0-\frac{i}{2}\left(a\tau_2\otimes\sigma_0+\alpha\tau_1\otimes\sigma_2\right)\,.
\end{eqnarray*} 
For the zeroth layer part of the Hamiltonian, we let $\tilde h=(h_A+h_B)/2$. We chose the parameters in the Hamiltonian to be $a=1$, $b=1$, $t=1.5$, $d=0.1$, $\alpha=0.3$, $\lambda=0.5$, and $\beta=0.4$. Fig.\ \ref{fig:ABstack}(c) can then be found by evaluating $A_{0,0}(\omega)$, which corresponds to the spectral density of the localized state at zeroth layer. Fig.\ \ref{fig:ABstack}(a) (Fig.\ \ref{fig:ABstack}(b)) corresponds to the spectral density of the surface states of the system A (B) by itself, which is obtained by turning off the contribution from the system B (A) by setting $h_B (h_A)=0$, $v_B(v_A)=0$ and $\tilde h=h_A (h_B)$ in evaluating $A_{0,0}(\omega)$. For Fig.\ \ref{fig:ABstack}(d), we calculated the spectral densities of the surface state of the system A for various values of $\omega$.

\subsection*{Conclusion}
We have explained how Weyl points act as monopoles for 2D FKMI, leading to a $\ZZ_2^4\oplus\ZZ$ classification of $\Twsm$ with four Weyl points, and generalised the analysis to $\Twsm$ with an arbitrary number of pairs of Weyl points. The Weyl points' history provides a mathematically dual classification in terms of arcs, which conveniently prefigures the Fermi surface on the surface. By ``rewiring'' Fermi arcs, Dirac cones can be produced/destroyed without a topological phase transition. Finally, we have numerically confirmed our theoretical prediction that Fermi arcs at a $\Twsm$ interface can combine to produce an extra Dirac cone.

\section*{Acknowledgements}
G.C.T.~was supported by ARC grant DE170100149, K.S.~by Grants-In-Aid for Scientific Research (Grant No.~15K13531) from the MEXT, Japan, and K.G.~by JSPS KAKENHI Grant No.~JP15K04871. G.C.T also thanks AIMR, Tohoku University, and the Simons Center for Geometry and Physics, for their hospitality.

\bibliographystyle{elsarticle-num}
\bibliography{ref_tiwsm}

\end{document}